\documentclass{osa-article}
\journal{osajournal}

\articletype{Research Article}

\begin{document}

\title{Angular resolved light scattering from micron-sized colloidal assemblies}

\author{Pavel Yazhgur,\authormark{1,$^+$} Geoffroy J. Aubry,\authormark{1}  Luis S. Froufe,\authormark{1} and Frank Scheffold\authormark{1,*}}

\address{\authormark{1} Department of Physics, University of Fribourg, 1700 Fribourg, Switzerland }
\email{\authormark{+}pavel.yazhgur@unifr.ch} 
\email{\authormark{*}frank.scheffold@unifr.ch} 
\homepage{https://www3.unifr.ch/phys/en} 


\begin{abstract*}
Disordered dielectrics with structural correlations on length scales comparable to visible light wavelengths exhibit complex optical properties. Such materials exist in nature, leading to beautiful structural non-iridescent color, and they are also increasingly used as building blocks for optical materials and coatings. In this article, we study the single-scattering properties of micron-sized, disordered colloidal assemblies. The aggregates act as structurally colored supraparticles or as building blocks for macroscopic photonic glasses. We present experimental data for the differential scattering and transport cross-section. We show how we can adapt existing macroscopic models to describe the scattering from small colloidal assemblies outside the weak-scattering limit and entering the Lorentz-Mie regime. 
\end{abstract*}


\section{Introduction}
Colloidal crystals and disordered, correlated suspensions or particle packings can exhibit full or partial photonic bandgaps  \cite{Joannopoulos2008,Yablonovitch1987,John1987,liew2011short,he2020colloidal,Florescu2009,haberko2020transition}. A stop-band has a finite spectral width and light of this color is strongly reflected, leading to a spectral response with a peak of reflectance at a resonance wavelength $\lambda_\text{max}$. While full photonic bandgap materials have drawn a lot of attention in the past \cite{Joannopoulos2008,John1987}, partial or incomplete band gaps are equally impressive, particularly for three-dimensional materials where the fabrication of full bandgap materials is still elusive. Partial photonic band gaps lead to structural color, an ubiquitous phenomenon, leading to bright and beautiful colors in bird feathers, plants, and insects \cite{vignolini2012pointillist, prum1998coherent,forster2010biomimetic,magkiriadou2012disordered,jacucci2020limitations,forster2010biomimetic}. The quality and usefulness of such photonic materials does not primarily rely on the bandgap-quality and strength but various parameters such as the thickness, absorption, multiple scattering, and the details of the scattering process in a structurally hierarchical material \cite{magkiriadou2012disordered,jacucci2020limitations}. Moreover it would be desirable if one could compartmentalize colloidal photonic structures in small building blocks or pigments that can be arranged at will, tuned in displays, or printed like inks.    
\newline Recently, many experimental studies addressed densely packed spherical colloidal aggregates, often referred to as 'photonic balls' (PBs). The aggregates consist of nanoparticles with a diameter around $\sim 300$nm produced by selective solvent evaporation or spray-drying. Their refractive index typically lies between $n= 1.4-1.6$ and the PBs are suspended in air or a solvent, such as water with an index $n_s$ \cite{wang2019structural,park2014full,braun2011colour,moon2004electrospray,wang2013recent,yi2003generation,kim2008optofluidic,ohnuki2020optical,Vogel10845}. The size of these aggregates can range from the micron to the millimeter scale. Previous studies' primary goal was to design PBs of targeted sizes, sometimes using specific core-shell nanoparticle architectures or incorporating absorbers.  The PBs are optimized to be bright, with pure color and their material properties are adapted for a particular purpose or application. Nearly all of the published work has focused on the material aspects of PBs, such as their fabrication and characterization. Modeling aspects are usually limited to discussing scaling laws derived from Bragg scattering off crystal lattices that predict the structural color peak-position $\lambda_\text{max}$. In contrast, we know little how PBs optical properties evolve, starting from the known Mie-scattering properties of the nanoparticles. This lack of quantitative modeling and lack of angular resolved experimental scattering data is especially surprising considering that the study of PBs represents an ideal playground to investigate questions about structural color formation. It is possible to finely tune the scattering strength and the amount of multiple scattering inside the PB by choosing an appropriate size and index mismatch.  Thus the study of photonic balls may provide insight into the optical physics of hierarchically structured materials based on well-defined, primary building blocks. 
\newline In this work we study experimentally, theoretically and numerically the angular resolved scattering from micron-sized spherical colloidal aggregates. Namely, we address a scattering regime where the index mismatch of the primary particle $m=n_\text{NP}/n_s$ is substantially larger than one. At the same time, we restrict our discussion to moderate index mismatch, such that resonant near field effects or the onset of non-classical transport regimes, such as a full bandgap formation or Anderson localization of light, can be safely neglected \cite{Froufe-Perez2016,Abrahams1979,Joannopoulos2008}. 

\section{Methods}\label{sec:Methods}

\subsection{Particle synthesis and fabrication of 'Photonic ball' aggregates}
We synthesize polystyrene nanoparticles (NPs) using standard surfactant-free polymerisation using 4-vinylbenzenesulfonate as an ionic co-monomer \cite{chonde1981emulsion}. The mean diameter of the spheres is $d_\text{NP}=348$~nm with a typical polydispersity of $5\%$, see Figure~\ref{fig:SLS348nm}. From these we fabricate photonic balls (PB) by a solvent-drying process. To this end we add about 20 $\mu$L aqueous NP-dispersion, at 1$\%$ volume fraction, to 1 mL of anhydrous decanol. We obtain a water-in oil emulsion, where the NPs remain in the aqueous phase, either by applying a vortex mixer at 2700 rpm for 20s or by passing the mixture through a narrow constriction using two syringes. Since water is slightly soluble in decanol the droplets rapidly shrink which leads to the formation of solid PBs. We concentrate and purify the PB-dispersions by centrifugation. We redisperse the PBs in isopropanol several times and in a final step we evaporate the  isopropanol and thus obtain PBs in purified water at 5mM concentration of KCl. The electrolyte KCl is added to screen double-layer repulsions between the PBs. We obtain the mean size and the particle size distribution (PSD) from image analysis based on scanning electron microscopy (SEM). 
The number density of polysterene nanoparticles $\rho_\text{NP}=\frac{n_\text{NP}}{V}$ in a PB-suspension can be calculated from the mass density obtained by drying and weighing a known volume of a PB dispersion. In turn the photonic ball number density $\rho_\text{PB}=n_\text{PB}/V$ and the number density of nanoparticles are linked by $d_\text{PB}^3\varphi= N d_\text{NP}^3$ where $\varphi$ denotes the volume fraction of nanoparticles in the PB-aggregate

\subsection{Optical measurements}

We record experimental differential cross-section data by static light scattering (SLS) using a commercial light scattering spectrometer-goniometer operating at a laser wavelength of $\lambda=660$nm covering scattering angles $\theta$ from $15^\circ$ to $150^\circ$ (LS Instruments, Fribourg, Switzerland). Toluene measurements serve as a reference to calibrate absolute scale data.  We collect further experimental data about the low angle scattering regime using holographic particle characterization with a commercial instrument (xSight, Spheryx Inc., New York, USA) \cite{lee2007characterizing}. 
\newline We extract data about the total transport cross-section $\sigma^\ast$ from measurements of the diffuse optical transmittance versus wavelength dependence $T(\lambda)$  of more concentrated PB suspensions. The dispersions are contained in a flat cuvette of thickness $L=2$mm and the calibrated transmittance is measured using an integrating sphere.
Our broadband light source is a supercontiuum laser (Fianium, NKT Photonics, Denmark) with a typical output of $1-2$ mW$/$nm. We record the transmitted power with a UV-VIS fiber spectrometer (Thorlabs, 200-1000 nm) with a wavelength resolution of $\Delta \lambda \simeq 2$nm. Experimentally, the optical transmittance through an optically dense ($L > 5 l^{\ast}$) slab is determined by the transport mean free path $\ell^\ast$ via
$T=  \frac{\ell^{\ast}+z_{0}}{L+2z_{0}}$, where  $L$ is the slab thickness and $z_{0}\simeq l^\ast$ is the extrapolation length from diffusion theory, for details see refs.~\cite{lemieux1998diffusing,kaplan1994diffuse}. 
By measuring $T$, we obtain $\ell^\ast$ and thus the transport cross-section
$\left<\sigma^\ast\right> / \left<N\right>= \left(\rho_\text{NP}\ell^\ast\right)^{-1}$, per nanoparticle. To this end we neglect positional correlations of the PB's, which is justified by the relatively low volume fraction occupied by PBs in dispersions ($  5-15 \%$), the PB's polydispersity and the large size $d_\text{PB} \gg \lambda$.  We have verified that, by taking into account hard-sphere type correlations \cite{Fraden1990,rojas2004photonic}, the results would differ by 5\% at most. 

\subsection{Numerical calculations}

\begin{figure*}
    \centering
    \includegraphics[width=0.8\linewidth]{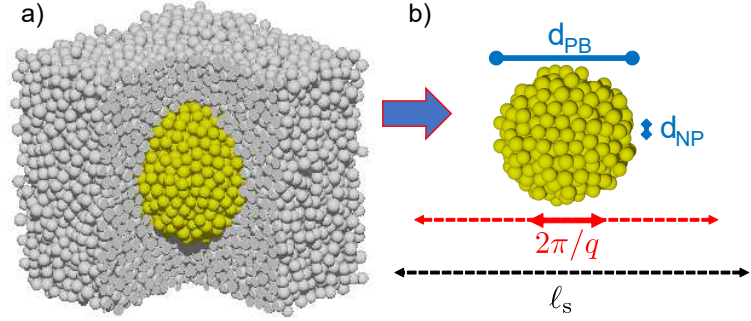}
    \caption{Molecular dynamics simulations of densely packed assemblies of nanoparticles with diameter $\text{d}_\text{NP}$. To mimic the experimental conditions we set the polydispersity to $\delta d/\overline d= 5$\% with $d_\text{NP}\equiv \overline d_\text{NP}$ for a filling fraction of $\varphi =0.60$. The photonic ball (PB) nanoparticle aggregates are cut from the bulk assembly and then used as input for the multiple-sphere T-matrix method (MSTM) to calculate numerically the differential scattering cross section of PBs. a) Large packings of about $N=$10000 nanospheres. b) Subset of $N$ particles contained in a sphere that defines the PB, $N$=443 in the example shown. Important length scales are $d_\text{PB}=\left (N/\varphi \right)^{1/3} d_\text{NP}$, the length scales probed in a scattering experiment $2\pi/q$ and the scattering mean free path $\ell_\text{s}\sim 15\mu$m for the polystyrene particle assemblies in water studied. We consider PBs with $d_\text{PB}\lesssim \ell_\text{s}$ or $N\lesssim 10 000$ and $d_\text{NP}=348$nm.}
    \label{fig:cuttingBall3}
\end{figure*}

We employ the multiple-sphere T-matrix method (MSTM open source code) to calculate numerically the differential scattering cross section of densely packed assemblies of nanoparticles (PBs)~\cite{Mackowski2011,MSTM}. The open-source code provides the exact time-harmonic electromagnetic scattering properties of the colloidal assembly.  The structures used as input for the MSTM code are generated by running a force-biased generation algorithm followed by a molecular dynamics equilibration using the PackingGeneration project~\cite{Baranau2014,PackingGeneration}. This algorithm supports polydisperse packings. To mimic the experimental conditions we chose to work with a polydispersity of 5\% with a filling fraction of $\varphi =0.60$. We first create large packings of about 10'000 nanospheres, and then select a subset of particles fully contained in a sphere, setting the diameter of the desired PB, Fig.\ref{fig:cuttingBall3}. A rendering of the structure is depicted in in Fig.~\ref{fig:cuttingBall3}~(a). Rather than using this sphere diameter directly, we define the PB-diameter by setting $d_\text{PB}^3\varphi= N d_\text{NP}^3$ to ensure that the mass of the PB is proportional to the aggregation number even for smaller $N$. As shown in Fig.~\ref{fig:Diameter} both measures for the sphere size deviate only slightly. The remaining difference is due to the influence of the rough surface. Using the MSTM code we perform calculations for PBs containing from $N=6$ to $1650$ nanoparticles. The numerical results presented in this work are obtained by averaging over all possible orientations of each PB which is directly supported in the MSTM package. 

\section{Results}
    
\subsection{Optical properties of bulk assemblies of nanoparticles}\label{subsec:bulk assemblies}
We start our analysis by considering a disordered bulk assembly of densely packed nanoparticles, in this context also called a photonic glass~\cite{Chen2017,rojas2004photonic}. Since we consider dense packings, the characteristic interparticle separation distance is dictated by the NP-size $d_\text{NP}$. The peak of the structure factor $S\left(q\right)$ for a disordered dense packing of hard spheres ($\varphi \simeq 0.6$) will lead to 
increased reflection for $\lambda_\text{max}/n_\text{s} \simeq 1.7 d_\text{NP}$, which is a vestige of Bragg-back scattering from Bragg planes, $\lambda_\text{max} \sim 2 d$, spaced at a distance $d$ \cite{NoteBragg,rojas2004photonic}. We note that this is a reasonable approximation only if the light is collected in the narrow range of angles around the backscattering direction. In general the position of the peak $\lambda_\text{max}$ slightly depends on the range of angles one uses to collect reflected light. Still, if we wish to engineer photonic materials in the visible range of wavelengths $\left (\lambda_\text{max}= 380-740 \text{nm}\right)$, we'll have to pack particles with a diameter of about $200-350$nm \cite{Vogel10845}.
\newline In this work, we study PBs that are sufficiently small such that random multiple scattering inside the PB can be neglected. We can easily derive the maximal size of a PB where this condition is met. To this end we first calculate the scattering mean free path $\ell_\text{s}$ of a photonic glass as discussed in the literature \cite{Fraden1990,Reufer2007}. PBs smaller than $\ell_s$ will be free of (internal) multiple scattering while PBs equal or larger $\ell_\text{s}$ will not.  The scattering mean free path for densely packed polystyrene particles with $n\simeq 1.59$, $\varphi \simeq 0.6$ of size $d_\text{NP}=348$nm is about $\ell_\text{s} \simeq 15 \mu$m in water 
for $\lambda =660$nm \cite{Fraden1990}. 

\subsection{Angular resolved single scattering properties} 
\begin{figure*}
    \centering
    \includegraphics[width=0.8\linewidth]{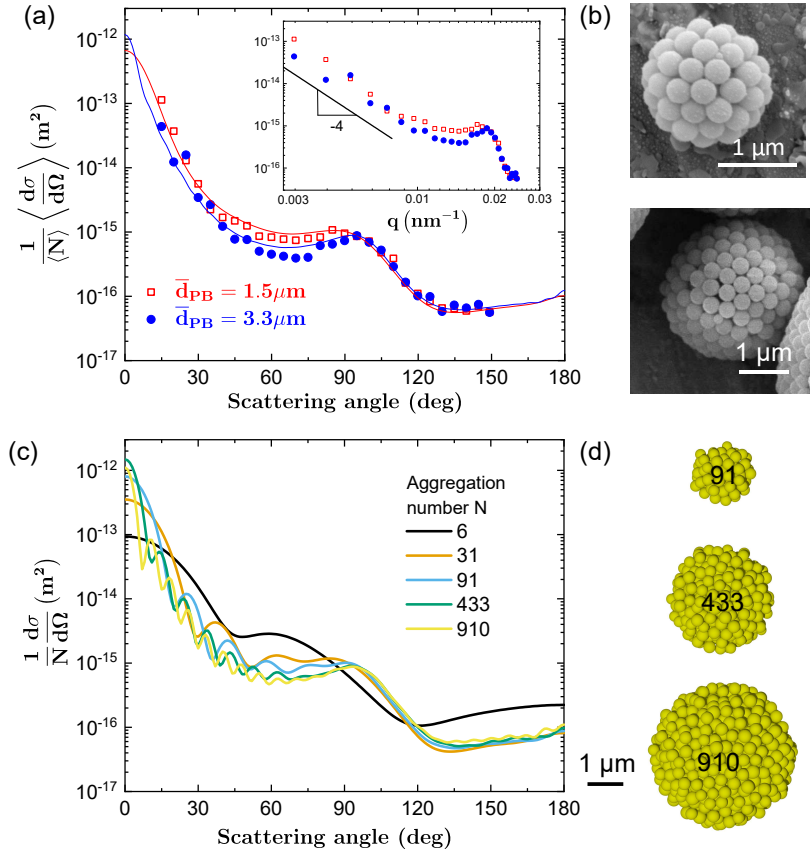}
    \caption{Static light scattering (SLS) from dilute suspensions of photonic balls suspended in water and composed of $d_\text{NP}= 348$nm polystyrene beads, $\lambda=660$nm.
    (a) Experimental differential scattering cross-section of fairly uniform PB suspensions. Open squares: PB with mean diameter  $\overline d_\text{PB}=1.5 \mu$m $\left(N\simeq 50\right)$ and polydispersity $\delta d_\text{PB}/\overline d_\text{PB} \sim 0.3$. Solid circles: PB's with a mean diameter  $\overline d_\text{PB}=3.3. \mu$m $\left(N\simeq 500\right)$ and polydispersity $\delta d_\text{PB}/\overline d_\text{PB} \sim 0.45$.   Solid lines show the predictions by MSTM-numerical calculations taking into account the PB-size distribution.   (b) SEM images of typical large and small PBs.
    (c) MSTM-differential scattering cross-sections for differently sized PBs. $N$ is the aggregation number of NPs and the corresponding PB-diameter $d_\text{PB}$ is given by $\varphi d_\text{PB}^3= N d_\text{NP}^3$ with $\varphi\simeq 0.6$. PB diameters from top to bottom are $d_\text{PB}\simeq0.75,1.30,1.86,3.12,4.00\ \mu$m.
    (d) Visualization of the PB's used in the simulations with $N=91, 433, 910$.}
    \label{fig:CSPB}
\end{figure*}

We discuss static light scattering from dilute, fairly uniform suspensions of photonic balls. 
We determine the differential scattering cross-sections of PB dispersions with the incident electric field polarized perpendicular to the scattering plane.
To compare the properties of differently sized PBs we divide the scattering cross-section by the mean aggregation number $N$ of PBs.  In Fig.~\ref{fig:CSPB}~(a) we show the experimental results for photonic balls with mean diameters $\overline d_\text{PB}= 1.5 \mu$m $\left(N\simeq 50\right)$, polydispersity $\delta d_\text{PB}/\overline d_\text{PB} \sim 0.3$ and $\overline d_\text{PB}=3.3 \mu$m $\left(N\simeq 500\right)$, polydisperstiy $\delta d_\text{PB}/\overline d_\text{PB} \sim 0.45$. The size of the NPs in both cases is $d_\text{NP}=348$~nm.
We also show SEM pictures of the PB particle aggregates in Fig.~\ref{fig:CSPB}~(b).
This choice of sizes was made to explore the scattering properties in two different limits. The smaller size only contains tens of NPs, while the larger PB already contains hundreds of NPs. The still relatively small overall size, $d_\text{PB}<5\mu$m for both cases, ensures that we can neglect multiple scattering inside the PBs. Interestingly, the scattering profiles recorded for the two differently sized PB do not differ by much. 
\newline The angular dependence of $\left<d\sigma/d \Omega\right> (\theta)$ at $\lambda=660$nm show two distinct regimes, Fig.~\ref{fig:CSPB}~(a). For $\theta <50^\circ$, the scattered intensity rises sharply towards smaller angles, which is a signature of the strong forward scattering of the PB-aggregates.
The log-log plot of the cross section, inset of Fig.~\ref{fig:CSPB}~(a), shows an approximate power-law scaling in momentum or $q$-space. This $q^{-4}$-scaling, in the limit $q>2\pi/R_\text{PB}\sim 3 \times 10^{-3} \text{nm}^{-1}$, resembles Porod's law for scattering from flat interfaces  \cite{glatter2018scattering}.  Here we recover the Porod-like envelope because the oscillations due to the single-sphere form-factor are smeared out owing to the PB size-polydispersity.
For larger angles, $\theta \geq 50^\circ$, the scattered intensity shows two shallow local minima and a peak at $\theta_\text{max} \simeq 95^\circ \pm 5^\circ$, which is associated to the short-range, liquid-like order, of the NPs inside the PB. The momentum transfer $q_\text{max}\simeq \left ( 4 \pi n_\text{s}/\lambda\right) \sin \left(\theta_\text{max}/2\right) \simeq 0.019 \text{nm}^{-1}$ matches the value expected for densely packed colloids  $q_\text{max}\simeq 2.3 \pi/d_\text{NP}=0.021 \text{nm}^{-1}$ \cite{liu2000improved}. $n_\text{s}=1.33$ denotes the solvent refractive index of water. The first minimum is a vestige of the suppressed long-range density fluctuations of the jammed NPs \cite{dreyfus2015diagnosing,torquato2015ensemble}, with a lower angle cut-off set by the PB-size. The second minimum is directly related to the form factor of the NPs \cite{bohren2008absorption}. 
The solid lines in Fig.~\ref{fig:CSPB}~(a) show that the numerical MSTM-results convoluted with the PSD perfectly match the experimental data for $\overline d_\text{PB}=1.5\ \mu$m and $\overline d_\text{PB}=3.3\ \mu$m.
\newline We note that, at a first glance, assuming a disordered assembly might seem in contradiction with the ordered structures observed in the SEM images, Fig.~\ref{fig:CSPB}~(c). It is known however \cite{Vogel10845}, that the inner parts of PB are typically much less crystalline than the outer shell of particles. We have verified that the internal structure of our PBs is indeed disordered by breaking the PBs with ultrasonification, as shown in Figure~\ref{fig:PBinside}.   Moreover, even in the presence of some remaining crystalline shells or domains, this will have a very limited impact on the scattering properties of small PBs. To illustrate the similarity of the $\left<d\sigma/d \Omega\right>/\left<N\right>$ between crystalline and disordered PBs of the same size, we include a comparison in Fig.~\ref{fig:CryDiso}. We note that crystallinity would play an increasing role for larger aggregates, i.e. larger than the so called Bragg length $L_B \sim l_\text{s}\sim 10 \mu$m, see refs.\cite{Vogel10845,spry1986theoretical}.
\newline In Fig.~\ref{fig:CSPB}~(c) we report detailed numerical results of $d\sigma/d\Omega~(\theta)$ for monodisperse PBs. Overall, spectra are very similar to the polydisperse case. Some resonant oscillations, that were washed out by polydispersity, appear at smaller angles.   
Our data suggest that at low angles, or $q d_\text{NP}\ll 2 \pi$, the optical properties of the PBs are governed by the scattering from the entire PB and the details of the internal structural properties are irrelevant, i.e. the PBs scatter as if they were homogeneous spheres.
To the contrary, for $q d_\text{NP} \sim 2.3 \pi$, we mainly probe the internal structure. Separating data in characteristic $q$-intervals is typical for scattering studies of hierarchically structured objects and scattering techniques are widely used to probe soft materials on different length scales \cite{lindner2002neutrons,genix2017determination}. 

\subsection{Small angle scattering from photonic balls} 
\begin{figure*}
    \centering
    \includegraphics[width=0.8\linewidth]{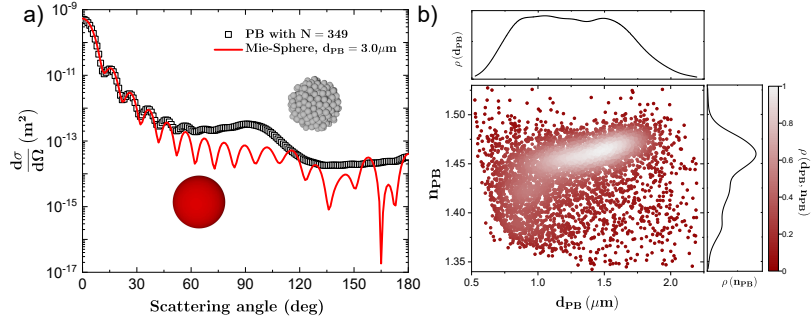}
    \caption{Small angle scattering properties of photonic balls. a) differential scattering cross section of PBs in water, $n_s=1.33$ per nanoparticle at $\lambda=660$nm. Open squares: numerical calculations with $N=349$ ($d_\text{PB}\simeq 2.9\mu$m ), nanoparticles of size $d_\text{NP}=348$nm and refractive index $n_\text{NP}=1.59$ for polystyrene. Red solid line shows the best small angle fit ($\theta<40
   ^\circ$) with Lorentz-Mie theory for a homogeneous sphere. We obtain a diameter $d_\text{PB}=2.98\mu$m and refractive index $n_\text{PB}=1.484$. b) Holographic particle characterization of PBs in water. The sample was prepared using the same experimental conditions as the smaller PBs discussed in Fig.~\ref{fig:CSPB} but taken from a different batch. Laser wavelength $\lambda=447$nm. 
   Joint probability distribution of $\rho(d_\text{PB},n_\text{PB})$ with mode $\overline d_\text{PB}=1.49 \pm 0.35\mu$m and  $\overline n_\text{PB}=1.46 \pm 0.03$.}
    \label{fig:hSphereRGD}
\end{figure*}

To examine the different contributions to the scattering cross section $\frac{1}{N}\frac{d\sigma}{d \Omega}(\theta)$, we start by modelling the small angle scattering properties. 
Since angles below $\theta \simeq 15^\circ$ are not accessible to our SLS-experiment we initially focus our attention on the comparison to the numerical data. In Figure \ref{fig:hSphereRGD}~(a) we show a Lorentz-Mie fit to the numerical results at low angles obtained for a $d_\text{PB} \approx 2.9 \mu$m photonic ball \cite{bohren2008absorption}. We find excellent agreement for all angles $\theta <40^\circ$ by adjusting the effective index to $ n_\text{PB}=1.484$ and the diameter $d_\text{PB}=2.98 \mu$m.
The fitted value for $n_\text{PB}$  closely matches the Maxwell-Garnett~\cite{Garnett1904} effective refractive index $n_\text{MG}=n_\text{s} \sqrt{\frac{(m^{2}+2)+2 \varphi (m^{2}-1)}{(m^{2}+2)- \varphi (m^{2}-1)}}\simeq1.483$, where $m=\frac{n_\text{NP}}{n\text{s}}\simeq 1.195$ denotes the refractive index contrast between the polystyrene nanoparticles and the aqueous solvent. 
\newline We collect experimental data about the low angle scattering regime using holographic particle characterization \cite{lee2007characterizing} with a commercial instrument (xSight, Spheryx Inc., New York, USA). The instrument analyzes individual particles delivered to a microscope's focal plane via a microfludic chip. The incident collimated laser beam $\left(\lambda=447\text{nm} \right)$ interferes with light scattered from the PBs. The instrument records holograms using a high numerical aperture oil-immersion objective (NA$=1.4$), thus probing a range of scattering angles $\theta \in \left [0^\circ,70^\circ\right]$, for details see ref.~\cite{odete2020role}. From the Lorentz-Mie fit of the holograms the instrument provides the particle size $d_\text{PB}$ and refractive index $n_\text{PB}$ for individual PBs. Single particle measurements are combined to population distributions as shown in Fig.~\ref{fig:hSphereRGD}~(b). 
The mean refractive index averaged over the several thousands of particles analyzed is $\overline n_\text{PB}=1.46\approx n_\text{MG}$. Our findings confirm previous studies on nano-porous silica particles, as well as small nanoparticle and protein-aggregates~\cite{odete2020role}. Moreover, we find that the effective medium approach remains valid even for NPs with sizes comparable to the wavelength (Mie-scatteres) with a relatively high refractive index contrast $m\simeq 1.2$.   
We conclude that, based on numerical and experimental data, the scattering from PBs in the limit $q d_\text{Np}\ll 2 \pi$ is described quantitatively by Lorentz-Mie theory, when assuming scattering from a homogeneous sphere with an effective refractive index  $\overline n_\text{PB} \simeq n_\text{MG}$. 

\subsection{Analytical model for angular dependent scattering from colloidal assemblies} Bulk crystalline or disordered assemblies of nanoparticles are translationally invariant. This invariance can be used to derive analytical expression for the scattered and interfering waves in the limit of weak scattering. In crystals, this leads to the concepts of Bloch waves and Bragg-scattering \cite{kittel1996introduction} while in liquids it allows us to define the isotropic structure factor $S\left(q\right)$ \cite{lindner2002neutrons}. In this section we aim to connect the scattering from finite sized PBs, with correlated disorder, to their macroscopic, translationally invariant, counterparts. Moreover we address the effect of corrections that arise for higher refractive index contrast, i.e. when the condition for weak-scattering are not met. We derive approximate analytical expressions for the differential scattering cross-section $d \sigma/d\Omega \left ( \theta \right)$. 
\newline We first consider a homogeneous assembly of identical spheres.
In the Rayleigh-Gans-Debye (RGD) approximation, also known as the 1\textsuperscript{st} Born approximation,  the effective differential scattering cross section of a nanoparticle assembly is given by the product of the particle's RGD cross section and the structure factor~\cite{glatter2018scattering},
\begin{equation}
\frac{1}{N} \frac{d\sigma}{d\Omega}\left(\theta,\lambda\right)=S\left(q \right ) \frac{d\sigma}{d\Omega}_\text{NP}\left(\theta,\lambda\right).\label{Eq:CSA}
\end{equation}
In this weak scattering limit, the scattering angle $\theta$ and the wavelength $\lambda$ are coupled such that the differential scattering cross-section can be expressed entirely in terms of the momentum transfer $q=2 k \sin \left (\theta/2\right)$. Eq.~\eqref{Eq:CSA} is commonly used when analyzing scattering experiments using X-rays, neutrons or light \cite{lindner2002neutrons}. The refractive index of the background medium $n_\text{s}$ enters for the wavenumber in the medium $k=2\pi n_s/\lambda$ \cite{Fraden1990}. The diffuse light propagation in a dense colloidal suspension or photonic glass can be considered as a train of uncorrelated single scattering events. More generally, this concept is known under the name diffuson- or ladder-approximation, for details see ref.~\cite{akkermans2007mesoscopic}. For randomly positioned point-like scatterers, $S(q)\equiv 1$ and the diffusion-approximation holds for $k l_s\gg1$ or $l_s\gg\lambda$.
For finite-size scatterers, with center positions that are spatially correlated but translationally invariant, the diffuson-approximation holds whenever the mean free path $\ell_\text{s}$ is both large compared to the structural correlation length $\xi$ \cite{kaplan1994diffuse} and the wavelength \cite{Leseur2016}. 
When using Eq.~\eqref{Eq:CSA} to calculate the single scattering function, this approach is also known as the collective scattering approximation (CSA) \cite{Fraden1990,ashcroft1966structure}. 
It is important to note that the CSA is much more restrictive than the diffuson-approximation. The latter only requires $l_s\gg\lambda,\xi$ which may hold even for large $m$, well beyond the validity of Eq.~\eqref{Eq:CSA}.
As a matter of fact, the range of validity of the CSA is very limited. To this end, already in the original work  \cite{Fraden1990}, Fraden and Maret had to modify the CSA and replace $\frac{d\sigma}{d\Omega}_\text{NP}\left(\theta,\lambda\right)$ in Eq.~\eqref{Eq:CSA} ad-hoc by the Lorentz-Mie scattering function.
It is only under this assumption that Eq.~\eqref{Eq:CSA} describes the diffuse scattering of fairly high index aqueous polystyrene particle NP-suspensions, $m\simeq 1.2$, up to volume fractions of 50\%  \cite{Fraden1990,kaplan1994diffuse, rojas2004photonic}. Although this hybrid-approach is common and seems plausible, it cannot be derived from Eq.~\eqref{Eq:CSA}.
For an exact theory, a Mie-type solution is needed. Moreover, in later work on high index particles $m> 2$ the host medium index $n_s$ has been replaced by an effective refractive index $n_\text{eff}>n_\text{s}$ in $S(q(\theta))$\cite{Reufer2007,Aubry2020,aubry2017resonant,busch1995transport}. 
\newline With the above mentioned modifications, Eq.~\eqref{Eq:CSA} has been quite successful for modeling the scattering and transport of light in optically dense suspensions and particle packings \cite{Fraden1990,kaplan1994diffuse,rojas2004photonic}. The characteristic scattering and transport mean free path $\ell_\text{s}=\left(\sigma \rho_\text{NP}\right)^{-1}$ and $\ell^\ast=\left(\sigma^\ast \rho_\text{NP}\right)^{-1}$ can be derived from the total scattering cross sections $\sigma,\sigma^\ast$ by integrating~Eq.~\eqref{Eq:CSA} over all solid angles $\text{d}\Omega$. However, a study of angular resolved scattering  $\frac{d\sigma}{d\Omega}\left(\theta,\lambda\right)$ for $m\ge1.1$ has lacked due to the difficulties accessing such highly turbid samples.
\newline To overcome this limitations, we study finite sized colloidal assemblies with $d_\text{PB}\le 10 \mu$m where we can perform single scattering experiments in large cuvettes containing only a small amount of PBs. We describe scattering from PBs by  replacing the bulk structure factor $S\left(q\right)$ in Eq.~\eqref{Eq:CSA} with the photonic ball structure factor $S_\text{PB}\left(q\right)$. The latter is defined as the finite sum of spherical waves emitted from NP-particle centers inside the PB (vectors in bold face):
 \begin{equation}
 S_\text{PB} (q) =\frac{1}{N} \left < \sum_{p,p'}e^{i\boldsymbol{q}\cdot\left(\boldsymbol{r}_{p}-\boldsymbol{r}_{p'}\right)}\right >.
 \label{eq:PBstructurefactor1}
\end{equation}  Eq.~\eqref{eq:PBstructurefactor1} can be rewritten for polydisperse suspensions or packings in order to obtain the RGD-measurable PB structure factor
\begin{equation}
 S^\text{m}_\text{PB}(q)= \left < \frac{\sum_{p,p'} V_\text{NP}^{p} V_\text{NP}^{p'} F(\boldsymbol{q}, R_\text{NP}^{p})F^{*}(\boldsymbol{q}, R_\text{NP}^{p'} ) e^{i\boldsymbol{q}\cdot\left(\boldsymbol{r}_{p}-\boldsymbol{r}_{p'}\right)}}{\sum_{p} \left| V_\text{NP}^{p} F(\boldsymbol{q}, R_\text{NP}^{p})\right|^{2}} \right >,
  \label{eq:PBstructurefactorpolydisperse}
\end{equation}
where $V_\mathrm{NP}^p$ and $F(\boldsymbol{q}, R_\text{NP}^{p})$ are the volume and scattering amplitude of nanoparticle $p$, respectively. Here we explicitly assume a linear superposition of the scattering from NPs  located at different positions. We neglect higher order scattering contributions of the Mie-type. Therefore, by definition, Eq. \eqref{eq:PBstructurefactorpolydisperse} cannot predict the Mie-scattering contributions, in particular the behaviour found at small angles shown in Fig.~\ref{fig:hSphereRGD}.
Still, we can use the Mie-scattering amplitudes of the NPs and slightly improve the weight given to the different components and then evaluate Eq. \eqref{eq:PBstructurefactorpolydisperse} numerically. In practice, however, we find that the difference is negligible, see supplementary Fig. \ref{fig:MieAmplitudes}.   
\newline Assuming we know the bulk structure factor, we can construct an approximate analytical theory to predict scattering from 
photonic balls, without the need to explicitly calculate the sum in Eq.~\eqref{eq:PBstructurefactorpolydisperse}. For clarity, we summarize the main steps, while the detailed derivation can be found in the Supplemental Methods.  For the NP cross section we can either use the RGD-expression, $\frac{d\sigma}{d\Omega}_\text{NP} \left(q\right)$, which makes the theory exact in the limit $n_\text{NP}\to n_\text{s}$, or we use the Lorentz-Mie result for $\frac{d\sigma}{d\Omega}_\text{NP}\left(\theta,\lambda\right)$ which provides a better approximation for higher index contrast but still neglects higher order scattering, near field coupling of the NPs in contact, and effective medium contributions. After some algebra, we can write the photonic ball structure factor as a convolution of the measurable structure factor of a polydisperse suspension of NPs $S^\text{m}(\boldsymbol{q})$ and the RGD monodisperse photonic ball form factor $P\left(\boldsymbol{q},  R_\text{PB}\right)$,
\begin{equation}
 S^m_\text{PB} (q) =V_\text{PB} \int S^\text{m}(\boldsymbol{q'})   P\left(\boldsymbol{q}-\boldsymbol{q'},  R_\text{PB}\right) d \boldsymbol{q'}.
 \label{Eq:PB_convolution1}
 \end{equation}
 
\begin{figure*}
    \centering
    \includegraphics[width=.8\linewidth]{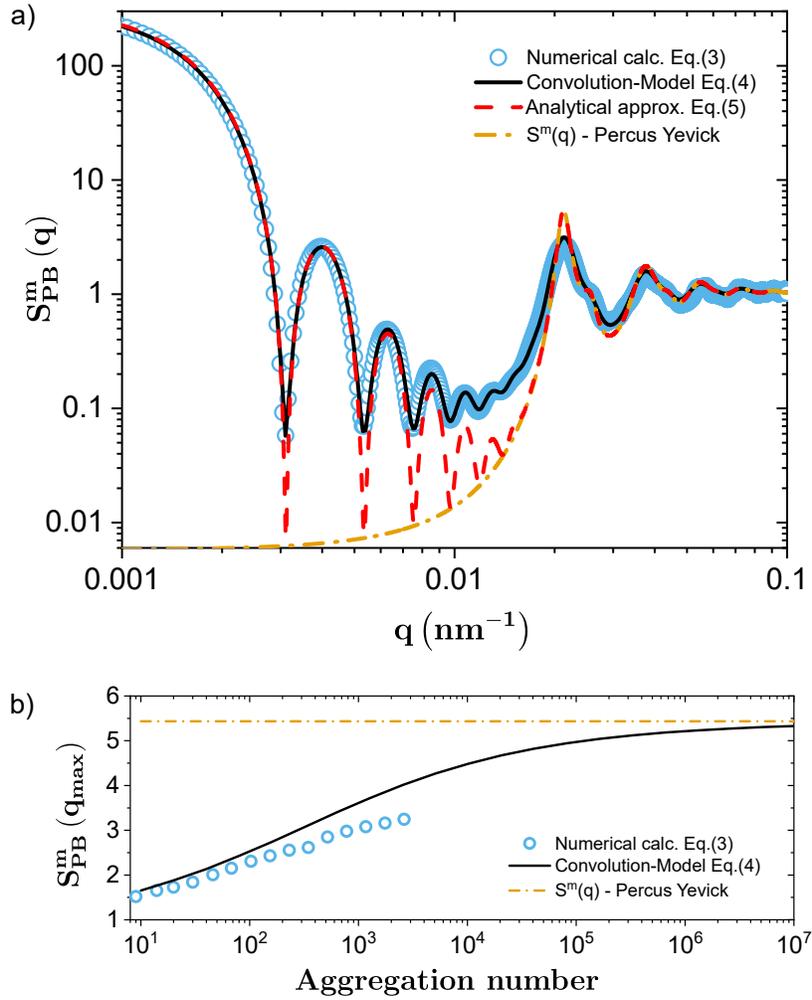}
    \caption{a) Photonic ball measurable structure factor for $N=349$, polydispersity 5\%, calculated from the convolution model, Eq.~\eqref{Eq:PB_convolution1}, the simple analytical model, Eq.~(\ref{Eq:PBCSA1}), and by the direct calculation from a simulated configuration, as shown in Fig. \ref{fig:CSPB}, using  Eq.~(\ref{eq:PBstructurefactorpolydisperse}). The measureable structure factor $S^\text{m}(q)$ of a corresponding Percus-Yevick fluid at $\varphi=0.6$ is also shown. b) The peak height at $q_\text{max}$ as a function of $N$ for the same model assumptions as in a). The data for Percus-Yevick fluid and the simple analytical approximation, Eq.~(\ref{Eq:PBCSA1}), are indistinguishable.  }
    \label{fig:AnalyticalModel}
\end{figure*}

\begin{figure*}
\centering
    \includegraphics[width=0.8\linewidth]{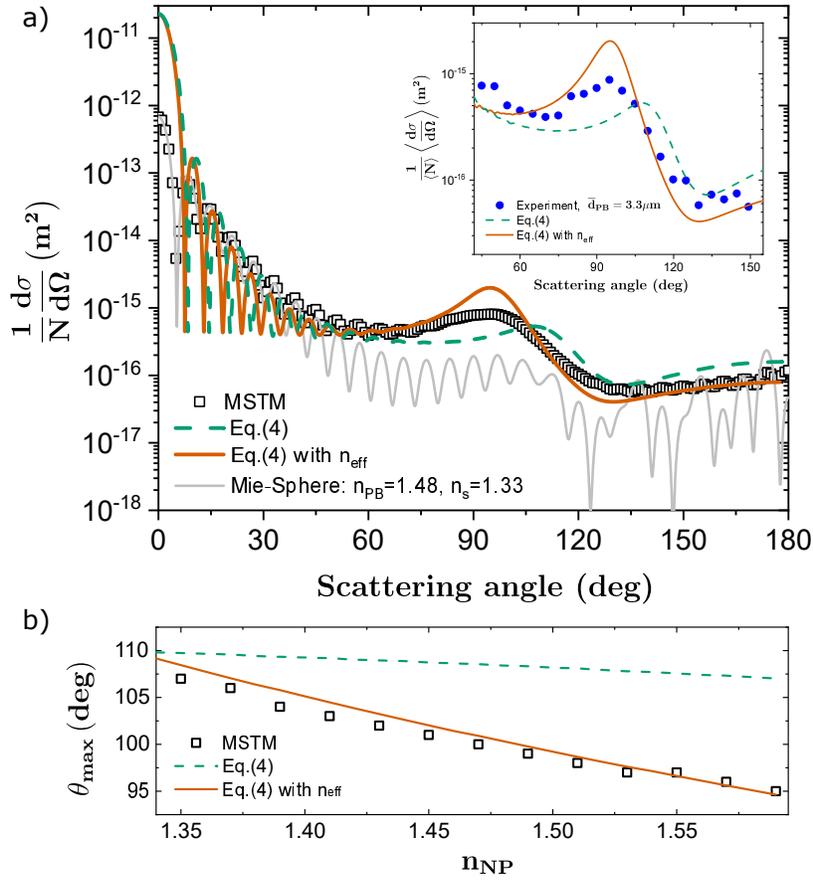}
        \caption{Comparison of the analytical model for the PB scattering function with the numerical (MSTM) differential scattering cross section. a) MSTM cross-section of a PB (open squares) with $N=1650$ ($d_\text{PB}\simeq 4.9\mu$m ),  $\lambda=660$nm, $n_\text{NP}=1.59$ and $n_\text{s}=1.33$. Grey solid line: Lorentz-Mie prediction for a homogeneous sphere, diameter $d_\text{PB}\simeq 4.8\mu$m. Green dash line: Convolution model Eq.~\eqref{Eq:PB_convolution1} based on the measurable Percus-Yevick structure factor and polydisperse Mie NP form factor, $q=2 k \sin{\left (\theta/2\right)}$ and $k=2 \pi n_\text{s}/\lambda$. Orange line: same model but replacing $q$ by $q_\text{eff}=q \frac{n_\text{MG}}{n_\text{s}}$ in the structure factor. The effective media refractive index $n_\text{eff}=1.483$ is calculated using the Maxwell Garnett approximation. Inset: Comparison of experimental differential scattering cross-section of PB dispersion with a mean diameter  $\overline d_\text{PB}=3.3 \mu$m and polydispersity $\delta d_\text{PB}/\overline d_\text{PB} \sim 0.45$ (blue circles) and the same analytical model  but now convoluted with PB size distribution. b) $\theta_{max}$ from MSTM as a function of the nano particle refractive index $n_\text{NP}\left(\in{1.34,159}\right)$.  Solid line: model prediction using $q_\text{eff}=2 k_\text{MG} \sin \left( \theta/2\right)$. Dashed line: model prediction using the bare $q-$value.
        }\label{fig:PBana}
    \end{figure*}

Separating small and large length scales, we can derive an approximate analytical expression 
 \begin{equation}
  S^m_\text{PB} (q) \approx  S^m\left(q, R_\text{NP}\right)+N P(q,R_\text{PB}) \label{Eq:PBCSA1}
\end{equation} from Eq.~\eqref{Eq:PB_convolution1}, for details see Supplemental Methods. Scattering from a PB can thus be understood as the arithmetical sum of the scattering from the internal structure and the scattering of the entire homogeneous PB plus a cross-over term, neglected in Eq.~\eqref{Eq:PBCSA1}. 
\newline For a comparison of the analytical and the numerical predictions, based on the polydisperese packing shown in Fig.~\ref{fig:CSPB}~c), we use the analytic results by Ginoza et al. for the measurable Percus-Yevick structure factor $S^\text{m}(q)$ for Schultz-distributed polydisperse hard spheres ~\cite{ginoza1999measurable}, polydispersity $\delta d_\text{NP}/\overline d_\text{NP}\simeq 5\%$~\cite{scheffold2009scattering,frenkel1986structure}. It was shown earlier that for somewhat polydisperse packings, this analytical model for $S^\text{m}(q)$ remains a good approximation, even at very high particle densities $\varphi$ up to random close packing~\cite{scheffold2009scattering,frenkel1986structure}. 
Figure~\ref{fig:AnalyticalModel}(a) shows a comparison between the approximate expression, Eq.~\eqref{Eq:PBCSA1}, the convolution model Eq.~\eqref{Eq:PB_convolution1}, and the direct numerical calculation of $S_\text{PB}\left (q_\text{max}\right)$ using Eq.~\eqref{eq:PBstructurefactorpolydisperse} based on the NPs positions obtained from the packing simulations. We find good agreement between all three curves at low angles ($q R_\text{RB}\to 0$) and large angles ($q R_\text{RB} \gg 1$). As expected, Eq.~\eqref{Eq:PBCSA1} deviates in the cross-over $q$-range, which corresponds to distances in between $R_\text{PB}$ and $R_\text{NP}$. 
\newline An important finding of this study is that the  local maximum of  $S^m_\text{PB} (q_\text{max})$ around $q\sim \pi/R_\text{NP}$ is reduced due to PBs finite size, expressed by the convolution with $P_\text{PB}(q)$, a fact that is clearly noticeable in Figure~\ref{fig:AnalyticalModel}(a) when comparing the prediction of Eq.~\eqref{Eq:PB_convolution1} and Eq.~\eqref{Eq:PBCSA1}. For the small PB-size ($N=349$) the peak value drops from more than 5 to about 3, almost by half, as shown in Figure~\ref{fig:AnalyticalModel}~b). The numerical value obtained using  Eq.~\eqref{eq:PBstructurefactorpolydisperse}, is even slightly lower compared to the prediction by Eq.~\eqref{Eq:PB_convolution1}, since the Percus-Yevick approximation, used in both Eq.~\eqref{Eq:PB_convolution1} and Eq.~\eqref{Eq:PBCSA1}, overestimates   $S^m_\text{PB} (q_\text{max})$ \cite{frenkel1986structure}. This finite-sized induced significant drop in  $S^m_\text{PB} (q_\text{max})$, summarized in Figure~\ref{fig:AnalyticalModel}(b),  adversely affects the photonic properties of small colloidal assemblies and their capabilities to be used as a colored pigment. 
\newline With $S^m_\text{PB}\left(q(\theta)\right)$ for a polydisperse aggregate, we can write the PB differential scattering cross section as follows,
\begin{equation}
 \frac{1}{N}\frac{d\sigma}{d\Omega}_\text{PB}\left(\theta,\lambda\right)=\frac{d\sigma \left(\theta,\lambda\right)}{d\Omega}_\text{NP} S^m_\text{PB} (q(\theta)),\label{Eq:CSA1}
\end{equation}
where $\frac{d\sigma \left(\theta,\lambda\right)}{d\Omega}_\text{NP}$ denotes the polydisperse NP cross section, which can be expressed analytically in the RGD-limit ~\cite{aragon1976theory} or calculated numerically using Mie-theory.
We note that using the approximation in Eq.~\eqref{Eq:PBCSA1} and taking the small $q(\theta)$ limit in Eq.~\eqref{Eq:CSA1}, we recover exactly the RGD-differential scattering cross section of a homogeneous sphere with a radius $R_\text{PB}$ and Maxwell-Garnett refractive index $n_\text{MG}$, see also ref.~\cite{odete2020role}.
\newline In Figure~\ref{fig:PBana} we compare the results from MSTM calculations to the predictions of Eq.~\eqref{Eq:CSA1} based on the full convolution model for $S^m\left(q\right)$, Eq.~\eqref{Eq:PB_convolution1}. We immediately notice that the data sets differ at low $q$. The difference at low $q \to 0$ is a direct consequence of the difference between our \emph{additive} model and Mie-scattering, which is not additive. Our model predicts that each NP contributes to the scattered field amplitude equally and thus at low $q \to 0$ $\frac{d\sigma}{d\Omega}_\text{PB} \sim N^2 \propto R_\text{PB}^{6}$. For Mie-scattering, the scattered amplitudes are reduced and eventually, for very large $(m-1) k R_\text{PB} \gg 1$, the forward scattered intensity scales with the square of the geometrical cross section $\frac{d\sigma}{d\Omega}_\text{PB}\sim R_\text{PB}^4\propto N^{4/3}$ \cite{hulst1981light}. We also observe differences around the local maximum  of $S_\text{PB}\left (q\right)$. The value of $q_\text{max}$ is not accurately predicted by the Eq.~\eqref{Eq:CSA1}. 
\newline Earlier studies suggested that for the relevant momentum transfer, probing structural correlations, one should consider an effective medium refractive index, therefore $k_\text{eff}=2 \pi n_\text{eff}/\lambda$\cite{busch1995transport,Reufer2007,magkiriadou2012disordered,Aubry2020}. Replacing $k$ by $k_\text{eff}\equiv k_\text{MG}$ for the definition of $q=2k\sin\left(\theta/2\right)$ in Eq.~\eqref{Eq:PB_convolution1} we find better agreement between the $\theta_\text{max}$ predicted by the model and the MSTM calculations, Figure~\ref{fig:PBana} (see also supplementary Figure~\ref{fig:PBanaSI}). In the inset of Figure~\ref{fig:PBana} we show the comparison to the experimental scattering data. 
We notice that while the peak position is predicted more accurately using $k_\text{eff}$, the absolute values around $q_\text{max}$ are overestimated. 
Using the slightly more accurate Eq.~\eqref{eq:PBstructurefactorpolydisperse}, instead of Eq.~\eqref{Eq:PB_convolution1}, does not lead to much agreement as shown in Fig. 6. The remaining discrepancy shows that replacing $k$ by $k_\text{eff}$  does not entirely overcome the limitations of our additive approach. 

\begin{figure}[h]
\centering
  \includegraphics[width=0.8\linewidth]{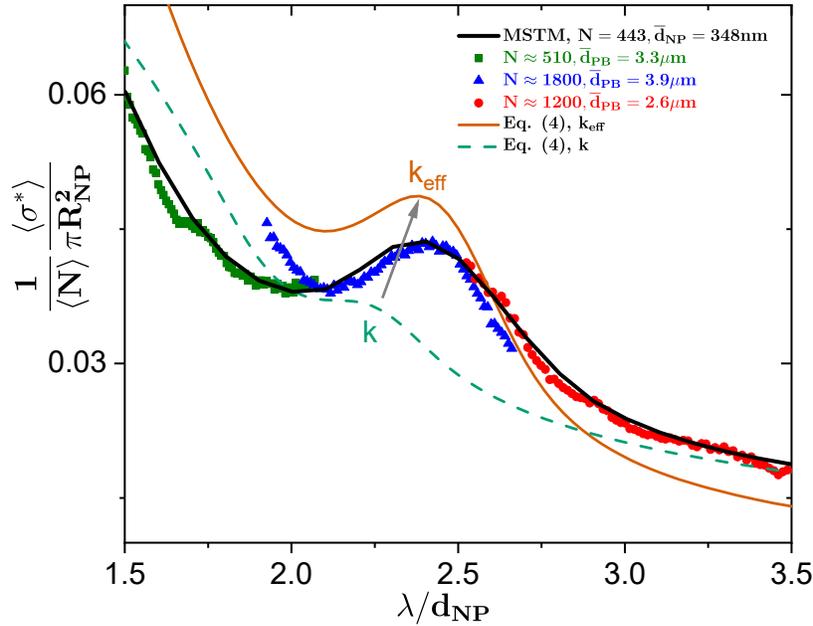}
  \caption{Transport cross-section as a function of wavelength for PBs suspended in water with different NP diameters: green squares $\overline d_\text{NP}=348$nm, $\overline d_\text{PB}=3.3 \mu$m ($N\approx 500$); blue squares $\overline d_\text{NP}=270$nm, $\overline d_\text{PB}=3.9 \mu$m ($N\approx 1800$); red squares $\overline d_\text{NP}=206$nm, $\overline d_\text{PB}=2.6 \mu$m ($N\approx 1200$). The $d_\text{NP}=348$nm data has been multiplied by a factor $0.9$.  The solid black line represent the MSTM-transport cross-section for a PB with N=443.  The colored lines display the transport cross-section derived from Eq.~\eqref{Eq:PB_convolution1} and Eq.~\eqref{Eq:CSA1}, using the measurable Percus-Yevick structure factor and the randomly polarized Mie-NP differential cross section with $n_{NP}=1.59$ and $n_{s}=1.33$:  solid line is 
  with Maxwell-Garnett (MG) effective media correction and dashed line without  
  }
  \label{fig:TransportCrossSection}
         \end{figure}

By integration $\sigma^\ast \left(\lambda\right)= 2 \pi \int_{0}^{\pi} \frac{d\sigma}{d\Omega}_\text{PB} \left(\theta,\lambda\right) \left(1-\cos\left(\theta\right)\right)\sin(\theta)  d\theta$ and taking the isotropic average over the polarization (randomly polarized incident light) we obtain the total transport cross-section $\sigma^\ast\left(\lambda\right)$ of a PB \cite{Fraden1990,rojas2004photonic}.
We find that For $R_\text{PB}\gg\lambda$ but  $R_\text{NP}\le\lambda$ the dimensionless cross section $\frac{\sigma^\ast}{N\pi R_\text{NP}^2}$ is predominately determined by the inner structure scattering of the PB and the dependence on the global size is weak, for details see supplementary Fig.~\ref{fig:TransportVSsize}. This weak dependence on the PB-size allows us to carry out experiments using different $d_\text{NP}$ and merge data from slightly different $R_\text{PB}$ to cover a larger range of $\lambda/d_\text{NP}$. In Figure \ref{fig:TransportCrossSection} we plot the experimental results for reduced $\sigma^\ast$ for three different NP-sizes. The experimental data covers a large wavelength-range and overall compares well to the MSTM prediction. For clarity we have rescaled the data for the largest NP sizes vertically by a factor $0.9$. We attribute the slight mismatch between experiment and MSTM-theory to PB-sedimentation, the remaining PB size-dependence as well as the limited accuracy when converting the measured total transmission data to $\left<\sigma^\ast\right>$, see also ref.~\cite{lemieux1998diffusing,kaplan1994diffuse}.  As expected, we find a local maximum at a characteristic reduced wavelength $\lambda_\text{max} / d_\text{NP} \sim 4 n_\text{eff}/2.3 \sim 2.5$. The convolution model properly predicts the overall shape of the curve over a quite wide range of wavelengths. Clear deviations are also visible. By using an effective refractive index in the structure factor calculations, the position of the peak can be reproduced, however the curve becomes tilted. This shows that neither using or $k$ or $k_\text{eff}$ leads to perfect agreement with the quantitative MSTM-predictions. 
 
\subsection{Extended simplified model taking into account the effective PB-scattering}
In the forward scattering direction, photonic balls behave as homogeneous effective Mie-spheres which we can describe quantitatively using an effective refractive index as shown in Fig.~\ref{fig:hSphereRGD}. In the opposite limit, for larger angles, Eq.~\eqref{Eq:PB_convolution1} and Eq.~\eqref{Eq:CSA1} provide good agreement but the additive model fails at small $q(\theta)$.
For the example shown in Fig.~\ref{fig:PBana}, the difference amounts to more than one order of magnitude for $q(\theta)\to 0$. Unfortunately there is no easy way to improve the accuracy of the full analytical model for $S^m_\text{PB} (q)$, Eq.~\eqref{Eq:PB_convolution1}.
It is however straightforward to improve the simplified expression, Eq.~\eqref{Eq:PBCSA1} together with Eq.~\eqref{Eq:CSA1}. To this end, we replace the second term with the exact Mie-scattering solution and obtain
\begin{equation}
\frac{1}{N}\frac{d\sigma}{d\Omega}_\text{PB}\left(\theta,\lambda\right)\approx \frac{d\sigma \left(\theta,\lambda\right)}{d\Omega}_\text{NP} S^m\left(q(\theta), R_\text{NP}\right )+\frac{1}{N}\frac{d\sigma \left(\theta,\lambda\right)}{d\Omega}_{\text{PB,Mie},n_\text{MG}}. \label{Eq:CSA1u}
\end{equation}

\begin{figure}[h]
\centering
  \includegraphics[width=1\linewidth]{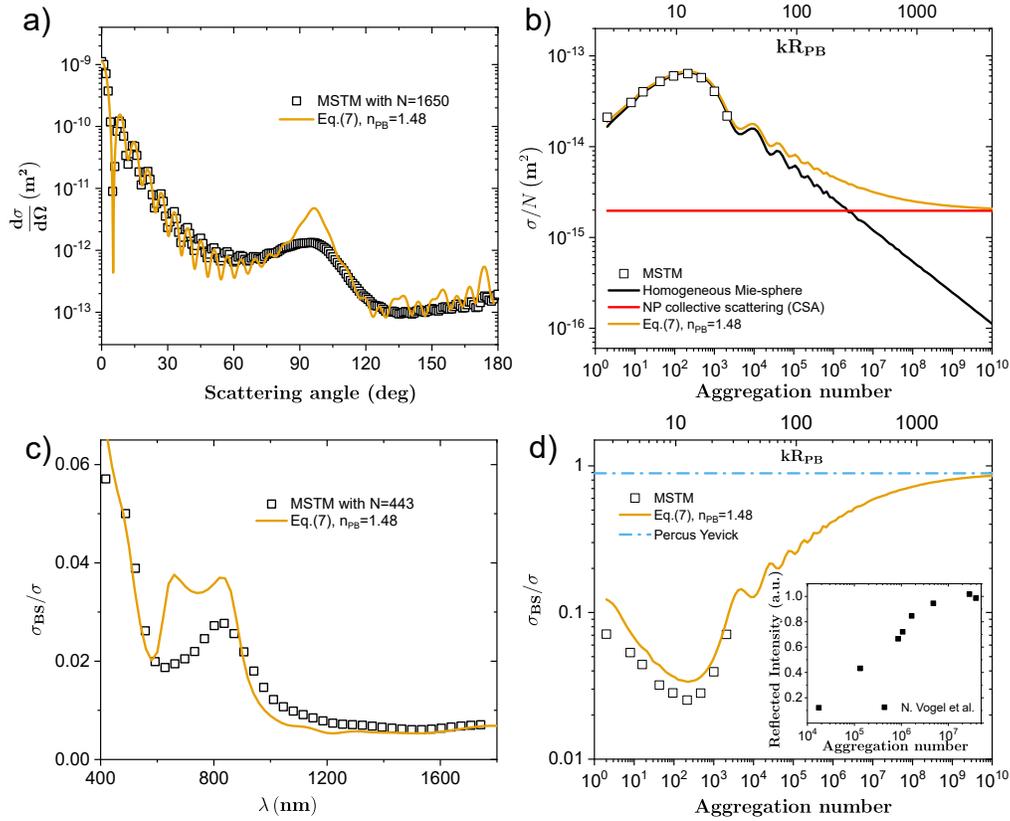}
  \caption{a) Extended simplified scattering model. Black open squares show the MSTM calculation of a photonic ball with $N=1650$ compared to the model prediction, Eq.~\eqref{Eq:CSA1u}. The nanoparticle refractive index is $n_\text{NP}=1.59$, the solvent index is $n_\text{s}=1.33$ and the mean index of the photonic ball is $n_\text{MG}=1.483$. b) Total scattering cross-sections for fixed wavelength $\lambda / d_\text{NP}=815/348=2.34$ and various aggregation numbers N. c)  Ratio of back-scattered to total scattering cross-sections as a function of wavelength for fixed $N=443$. d) Ratio of back-scattered to total scattering cross-sections for fixed wavelength $\lambda / d_\text{NP}=815/348=2.34$ and various aggregation numbers N. The blue solid line shows the corresponding prediction for an infinite photonic glass $\sigma_\text{BS}/\sigma=0.89$. The Inset: experimental data for the light intensity reflected  from similar size photonic balls consisting from polysterene NPs in air  (refractive index contrast $m \approx 1.59$), data reproduced from ref.\cite{Vogel10845}. The maximal intensity in the experiment is set to one. In all model predictions, we are using $q_\text{eff}$ for the measurable structure factor. 
  }
  \label{Fig:eCSA}
         \end{figure}

In Figure~\ref{Fig:eCSA}~(a) we show a comparison with MSTM data over the entire scattering angle range and find fairly good agreement, again using $q_\text{eff}=2k_\text{eff}\sin\left(\theta/2\right)$. As expected, the agreement is less good over a range of intermediate angles around the peak of the structure factor. Since we are using the bulk structure factor in Eq.~\eqref{Eq:CSA1u}, we overestimate the contribution of correlated scattering and peak height. Nevertheless, this simple model captures the small and large angle scattering limit accurately. Importantly, Eq.~\eqref{Eq:CSA1u} is based entirely on analytical results from Mie-theory and the measurable structure factor, readily accessible in the literature and straightforward to calculate \cite{bohren2008absorption,ginoza1999measurable}. 
\newline Using Eq.\eqref{Eq:CSA1u} we can, for example, understand the transition from a finite sized PB to a bulk assemblies of particles. We can integrate $Eq.\eqref{Eq:CSA1u}$ and express the total scattering cross-section as the sum of the NP-scattering and the effective sphere scattering $\sigma=N \sigma_\text{NP}+\sigma_{\text{PB,Mie},n_\text{MG}}$. In Figure ~\ref{Fig:eCSA}~(b) we plot both terms of $\sigma/N$ as a function of the aggregation number $N$. The NP-contribution is independent of $N$ and the second term shows a maximum around $N\sim 100$ ($kR_{PB}\sim 10$) related to the Mie resonance of the corresponding homogeneous Mie sphere with $n_\text{MG}$. For $R_\text{PB}\gg \lambda$ the effective sphere cross section  $\sigma_{\text{PB,Mie},n_\text{MG}}$ increases as the geometric cross section $R_\text{PB}^2\propto N^{2/3}$ and thus $\frac{\sigma_{\text{PB,Mie},n_\text{MG}}}{N}\propto N^{-1/3}$. The latter vanishes in the large $N-$limit and we recover the known result for a photonic glass \cite{Fraden1990,rojas2004photonic,Reufer2007,aubry2017resonant,Aubry2020}. Once more, we demonstrate that a sufficiently high aggregation number is necessary ($N \ge 10^6$)  to approach the infinite system behaviour.
\newline We can also use Eq.~\eqref{Eq:CSA1u} to estimate another important property of a photonic ball, namely its ability to preferentially reflect light at a resonant wavelength interval associated with the peak of the structure factor. Light that is not reflected will by scattered in other directions. In optically dense samples, composed of many PBs, scattered light will contribute to multiple scattering and the formation of a diffuse background. The latter can be suppressed by adding an absorber \cite{forster2010biomimetic,Schertel2019a} or by preparing thin films, but in either case the brightness is diminished because incident light of the desired color is not converted into reflected light.
\newline In Figure~\ref{Fig:eCSA}~(c) and (d) we plot $\sigma_{BS}$ of the light scattered at angles $\theta\in[90^\circ,180^\circ]$ divided by the total scattering cross section $\sigma$, with $\sigma_{BS}\left(\lambda\right)= 2 \pi \int_{\pi/2}^{\pi} \frac{d\sigma}{d\Omega}_\text{PB} \left(\theta,\lambda\right) sin(\theta)  d\theta$ and $\sigma \left(\lambda\right)= 2 \pi \int_{0}^{\pi} \frac{d\sigma}{d\Omega}_\text{PB} \left(\theta,\lambda\right) sin(\theta)  d\theta$. We show data obtained from MSTM and model predictions from Eq.~\eqref{Eq:CSA1u}. First we notice that for  $N\lesssim1000$  the ratio  $\sigma_{BS}/\sigma$  is surprisingly small over the entire spectral range, which shows that such small PBs predominately scatter in forward direction and thus are not well suited for structural colour applications. The photonic glass limit is only reached asymptotically for PBs with $N \ge 10^6-10^7$ or $d_\text{PB}>(\frac{N}{\varphi})^{1/3} d_\text{NP}\simeq 250 d_\text{NP}$ which in our case corresponds to $d_\text{PB}\gtrsim 40-100\mu$m. The solid lines show the model predictions which describe the data well but overestimate the amount of backscattering, in particular for small PB sizes. 
\newline Experimentally, Vogel et al. \cite{Vogel10845} observed a similar trend by measuring the intensity of light reflected from photonic balls consisting of polystyrene NPs in air ($m \approx 1.59$). As shown in the inset of Fig.~\ref{Fig:eCSA}~(d)), their reflection signal, obtained by illumination and detection through a microscope objective, enormously suffers from finite-size effects and at least qualitatively follows our scattering model prediction. This latter supports our claim that a quantitative scattering model can benefit photonic balls' design and characterization. 

 \section{Summary and Conclusion}   
 We have studied experimentally, numerically, and theoretically the light scattering properties of finite-sized hierarchical nanoparticle assembly with a refractive index mismatch clearly not covered by the weak scattering, or Rayleigh-Gans-Debye (RGD), theory. We could show that scattering from small aggregates, containing up to a few thousand NPs, can be modeled quantitatively using the multi-sphere-T-matrix (MSTM) formalism, where efficient algorithms are freely available. We have also shown that absolute scale experimental and theoretical light scattering data agree quantitatively. For larger assemblies, as well as for bulk structures, the T-matrix approach becomes computationally too expensive. Therefore, other numerical or approximate analytical methods are necessary to describe the experimental phenomena, such as structural coloration and parameters that control the total scattering and diffuse transport of light. Starting from scattering theory in the weak scattering limit, we have derived different simple approximations and ad-hoc improvements. These 
 models allow us to probe and verify certain assumptions independently. Importantly our studies show that an effective refractive index has to be used to reproduce the scattering peak $q_\text{max}$ or peak wavelength $\lambda_\text{max}$ associated with enhanced reflection at $\lambda_\text{max}$  for structural coloration.
 Our analytical model predictions compare well with the experiments and MSTM-data but cannot quantitatively describe the cross-section over the entire angular or spectral range. Our results also show that small PBs, containing less than $10^4$ nanoparticles, or smaller than $10 \mu$m, predominately scatter in the forward direction and thus produce little structural color. For such small PBs, we find that the reflected light amounts to less than 15\% of the total scattered light. Moreover, we find that for an index mismatch up to at least $m=1.2$, the forward scattering properties can be described by the Mie-theory for homogeneous spheres with an effective refractive index equal or close to the Maxwell-Garnett index. Based on these findings, we developed a simple extended collective-scattering approximation that captures both the small and large $q$-scattering properties accurately.


  \section{Acknowledgements}
 FS thanks the Center for Soft Matter Research (CSMR) at New York University for hosting him during his research sabbatical and for giving us access to the Spheryx xSight holographic particle sizer.  We are grateful to David Grier for discussions and help with the xSight data analysis. FS thanks Julian Oberdisse for illuminating comments concerning the RGD-modelling. PY thanks Chi Zhang for assistance with some of the optical measurements. The Swiss National Science Foundation financially supported this work through the National Center of Competence in Research Bio-Inspired Materials, No.~182881 and through projects No.188494 and No.~183651.

\bibliography{biblioPB1}

\clearpage
\newpage
\setcounter{equation}{0}
\setcounter{figure}{0}
\setcounter{table}{0}
\setcounter{page}{1}
\renewcommand{\theequation}{S\arabic{equation}}
\renewcommand{\thefigure}{S\arabic{figure}}
\section*{Supplementary Material}
\title{Angular resolved light scattering from micron-sized colloidal assemblies}

\author{Pavel Yazhgur,\authormark{1,$^+$} Geoffroy J. Aubry,\authormark{1}  Luis S. Froufe,\authormark{1} and Frank Scheffold\authormark{1,*}}

\address{\authormark{1} Department of Physics, University of Fribourg, 1700 Fribourg, Switzerland }
\email{\authormark{+}pavel.yazhgur@unifr.ch} 
\email{\authormark{*}frank.scheffold@unifr.ch} 
\homepage{https://www3.unifr.ch/phys/en} 


\section*{Supplemental Methods}

\paragraph{Definition of the photonic ball radius}
For the MSTM-calculations we cut spherical assemblies from dense packing in a rectangular simulation box terminated with periodic boundary conditions.  The size of the sphere is known as the Feret diameter corresponding to  the edge length of a cube circumscribed around the PB, such that the outer layer of nanoparticles is fully included. This definition agrees with the way we may obtain sizes from an electron micrograph (SEM). The average density however drops gradually to zero at radii $r\in[R_\text{PB}-R_\text{NP},R_\text{PB}]$ and thus we expect the effective size in scattering to be slightly smaller. 
Indeed, the fitted radius obtain by analysis of the low-angle scattering, as show in Fig. \ref{fig:CSPB}, matches $R_\text{PB}$ defined by the total mass of the PB: $R_\text{PB}^3\varphi= N R_\text{NP}^3$. This definition is also consistent with RGD-limit where for $q \to 0$ the differential scattering cross section is proportional to the total mass of the scattering object.

\paragraph{Scattering matrix and differential cross-sections}
We consider an incoming plane wave 
\begin{equation}
\boldsymbol{E}_{in}\left(\boldsymbol{r},\omega\right)=\boldsymbol{E}_{in}e^{i\left(\boldsymbol{k}_{in}\cdot\boldsymbol{r}-i\omega t\right)}\label{eq:sct_10}
\end{equation}
We define the $z$ axis such that $\boldsymbol{k}_{in}\parallel\boldsymbol{u}_{z}$
($\boldsymbol{u}_{x,y,z}$ being the unitary vectors along the coordinated
axes). For a given scattering direction (see sketch in Fig.\ref{fig:sketch_coords}) given
by the polar ($\theta$) and azimuthal ($\phi)$ angles, the scattering
plane is defined by the incoming and outgoing directions $\boldsymbol{u}_{z}$
and $\boldsymbol{u}_{r}$ respectively. Since the scattered electric
field is perpendicular to $\boldsymbol{u}_{r}$, we define two independent
components in-plane and perpendicular to the scattering plane given
by the unitary vectors\begin{subequations}
\begin{align*}
\boldsymbol{\hat{e}}_{\perp s} & \equiv-\boldsymbol{\hat{e}}_{\phi},\\
\boldsymbol{\hat{e}}_{\parallel s} & \equiv\boldsymbol{\hat{e}}_{\theta}.
\end{align*}
\label{eq:sct_20}\end{subequations}Hence the asymptotic scattered
field is 
\begin{equation}
\boldsymbol{E}_{s}\left(\boldsymbol{r}\right)=\frac{e^{ikr}}{-ikr}\left[E_{\perp s}\boldsymbol{\hat{e}}_{\perp s}+E_{\parallel s}\boldsymbol{\hat{e}}_{\parallel s}\right]\label{eq:sct_30}
\end{equation}
 for some amplitudes $E_{\perp s}$ and $E_{\parallel s}$. We drop
the frequency dependency for clarity.
\newline Analogously, the incoming field can be expressed in components parallel
and perpendicular to the scattering plane
\begin{equation}
\boldsymbol{E}_{in}=E_{\perp in}\hat{\boldsymbol{e}}_{\perp in}+E_{\parallel in}\hat{\boldsymbol{e}}_{\parallel in}\textrm{,}\label{eq:sct_40}
\end{equation}
where $\hat{\boldsymbol{e}}_{\perp in}=\boldsymbol{\hat{e}}_{\perp s}$
and $\hat{\boldsymbol{e}}_{\parallel in}=\boldsymbol{u}_{z}\times\hat{\boldsymbol{e}}_{\perp in}$
are unit vectors perpendicular and parallel to the scattering plane
and both are perpendicular to $\boldsymbol{k}_{in}\parallel\boldsymbol{u}_{z}$.

The incoming and scattered amplitudes are related by the scattering
matrix $S$ (equivalent to the definition given in \cite{bohren2008absorption})
\begin{equation}
\left(\begin{array}{c}
E_{\perp s}\\
E_{\parallel s}
\end{array}\right)=\left(\begin{array}{cc}
S_{1} & S_{4}\\
S_{3} & S_{2}
\end{array}\right)\left(\begin{array}{c}
E_{\perp in}\\
E_{\parallel in}
\end{array}\right).\label{eq:sct_50}
\end{equation}
\newline In the general case, all entries of the scattering matrix depend on
the angles (for a given orientation of the scatterer in the coordinate
system), and we have
\begin{equation}
\boldsymbol{E}_{s}\left(\boldsymbol{r}\right)=\frac{e^{ikr}}{-ikr}\left[\left(S_{1}\sin\phi+S_{4}\cos\phi\right)E_{\perp s}+\left(S_{2}\cos\phi+S_{3}\sin\phi\right)E_{\parallel s}\right].\label{eq:sct_60}
\end{equation}
\newline The differencial scattering cross section, defined as the power scattered per unit solid angle in the direction $(\theta,\phi)$ normalized to the input intensity is 
\begin{equation}
\frac{d\sigma\left(\theta,\phi\right)}{d\Omega}=\frac{\left|S_{1}\sin\phi+S_{4}\cos\phi\right|^{2}+\left|S_{2}\cos\phi+S_{3}\sin\phi\right|^{2}}{k^{2}}.\label{eq:sct_70}
\end{equation}
\newline Some particular cases are:
\newline\textbf{Scattering in the plane perpendicular to the polarization}.
We consider linear polarization along the $x$ axis, and the scattered
radiation is measured in the $z-y$ plane, i.e. $\phi=\pi/2$. The scattered
field is 
\begin{equation}
\left.\boldsymbol{E}_{s}\left(\boldsymbol{r}\right)\right|_{\phi=\pi/2}=\frac{e^{ikr}}{-ikr}\left[S_{1} E_{\perp s}+S_{3} E_{\parallel s}\right]\label{eq:sct_80}
\end{equation}
and the scattering cross section
\begin{equation}
\left.\frac{d\sigma\left(\theta,\phi\right)}{d\Omega}\right|_{\phi=\pi/2}=\frac{\left|S_{1}\right|^{2}+\left|S_{3}\right|^{2}}{k^{2}}\label{eq:sct_90}
\end{equation}
\newline\textbf{Scattering by an spherical object. }If the object shows spherical
symmetry and scalar isotropic refractive index, it can be shown that
the off-diagonal terms of the scattering matrix vanish. Hence
\begin{equation}
\frac{d\sigma\left(\theta,\phi\right)}{d\Omega}=\frac{\left|S_{1}\sin\phi\right|^{2}+\left|S_{2}\cos\phi\right|^{2}}{k^{2}}\text{,}\label{eq:sct_100}
\end{equation}
of course a particular case is the standard Mie theory. However this
form of the differential scattering cross section also holds for non-spherical
objects under certain approximations. Particularly in three relevant
cases:
\begin{itemize}
\item In the Rayleigh-Gans-Debye approximation the scattered field is considered
in first order approximation in scattering series (first order Born
approximation) and hence the scattered field is a superposition of
electric dipole fields that show the exact same symmetry. The relevant
point here is that the fields exciting the induced dipoles are aligned
with the input polarization. This symmetry holds irrespective of the
shape of the scattering object.
\item Assemblies of Mie scatterers in first order scattering approximation.
In this case the scattering by each individual building block is considered
exactly (Mie theory) and the multiple scattering between building
blocks is neglected. Hence the scattered field is a superposition
of the fields scattered by each building block and corresponds to
eq.(\ref{eq:sct_100}).
\item The averaged scattered field from objects that are spherical on average
(this should hold for instance when performing orientation average
of any object) also shows spherical symmetry,
\begin{equation}
\left\langle \boldsymbol{E}_{s}\left(\boldsymbol{r}\right)\right\rangle =\frac{e^{ikr}}{-ikr}\left[\left\langle S_{1}\right\rangle \sin\phi E_{\perp s}+\left\langle S_{2}\right\rangle \cos\phi E_{\parallel s}\right]\label{eq:sct_110}
\end{equation}
For each realization in the ensemble or particular orientation, we
can write the scattering matrix elements $S_{i}$ as
\begin{equation}
S_{i}=\left\langle S_{i}\right\rangle +\delta S_{i}.\label{eq:sct_120}
\end{equation}
If the object shows spherical symmetry on average, $\left\langle S_{3}\right\rangle =\left\langle S_{4}\right\rangle =0$
and the averaged differential scattering cross section takes the general
form
\begin{align}
k^{2}\left\langle \frac{d\sigma\left(\theta,\phi\right)}{d\Omega}\right\rangle = & \left|\left\langle S_{1}\right\rangle \sin\phi\right|^{2}+\left|\left\langle S_{2}\right\rangle \cos\phi\right|^{2}\nonumber \\
 & +\left\langle \left|\delta S_{1}\right|^{2}+\left|\delta S_{3}\right|^{2}\right\rangle \sin^{2}\phi+\left\langle \left|\delta S_{2}\right|^{2}+\left|\delta S_{4}\right|^{2}\right\rangle \cos^{2}\phi\nonumber \\
 & +2\text{Re}\left\{ \left\langle \delta S_{1}\delta S_{4}^{*}+\delta S_{2}\delta S_{3}^{*}\right\rangle \right\} \sin\phi\cos\phi .\label{eq:sct_130}
\end{align}
If we can neglect the fluctuation-correlation terms $\left\langle \delta S_{i}\delta S_{j}^{*}\right\rangle $ in Eq.~(\ref{eq:sct_130}), then $k^{2}\left\langle \frac{d\sigma\left(\theta,\phi\right)}{d\Omega}\right\rangle \simeq\left|\left\langle S_{1}\right\rangle \sin\phi\right|^{2}+\left|\left\langle S_{2}\right\rangle \cos\phi\right|^{2}$
and, in the $\phi=\pi/2$ scattering plane,
\begin{equation}
\left\langle \left.\frac{d\sigma\left(\theta,\phi\right)}{d\Omega}\right|_{\phi=\pi/2}\right\rangle \simeq\frac{\left|\left\langle S_{1}\right\rangle \right|^{2}}{k^{2}}\simeq\frac{\left\langle \left|S_{1}\right|^{2}\right\rangle}{k^{2}}\label{eq:sct_140}
\end{equation}

In particular, in Fig. \ref{fig:MieAmplitudes2} we demonstrate that off-diagonal elements of the scattering matrix can be safely neglected in the description of sufficiently small PBs.
\end{itemize}
In this last case, if incident light is randomly polarized, $\phi$ is distributed evenly.
We can then average over $\phi\in\left[0,2\pi\right]$ and find
\begin{equation}
\frac{d\sigma}{d\Omega}=\frac{\left|S_{1}\right|^{2}+\left|S_{2}\right|^{2}}{2k^{2}}\label{eq:sct_150}
\end{equation}
In this case, the total scattering cross section is given by
\begin{equation}
\sigma=\int_{0}^{2\pi}d\phi\int_{0}^{\pi}\frac{d\sigma}{d\Omega}\sin\theta d\theta=2\pi\int_{0}^{\pi}\frac{d\sigma}{d\Omega}\sin\theta d\theta\label{eq:sct_160}
\end{equation}
Another useful property is the average of the cosine of the scattering
angle given by
\begin{equation}
g:=\left\langle \cos\theta\right\rangle =\frac{\int_{0}^{\pi}\frac{d\sigma}{d\Omega}\cos\theta\sin\theta d\theta}{\int_{0}^{\pi}\frac{d\sigma}{d\Omega}\sin\theta d\theta}
\label{eq:sct_170}
\end{equation}
\newline The transport cross section $\sigma^{*}$ and the scattering cross section
can be linked using $\sigma^{*}=\left(1-g\right)\sigma$.
Using the
identity $\sin^{2}\left(\theta/2\right)=\left(1-\cos\theta\right)/2$, we
can write
\begin{equation}
\sigma^{*}=\frac{1}{k^{6}}\int_{0}^{2k}\frac{d\sigma}{d\Omega}q^{3}dq
\label{eq:sct_180}
\end{equation}
The momentum transfer is denoted as $q=2k\sin\left(\theta/2\right)$.
For moderately concentrated PB-suspensions up to about 10-15\% in volume fraction, the scattering and transport mean free paths are $\ell_\text{s} \simeq\left (\rho_\text{PB}\sigma \right)^{-1}$ and $\ell^\ast\simeq\left (\rho_\text{PB}\sigma^\ast \right)^{-1}$.

\paragraph{Convolution model for $S_\text{PB}(q)$}
To obtain an analytical expression for the PB structure factor from Eq. \eqref{eq:PBstructurefactor1}, we need to obtain the ensemble average of $\sum_{p,p'}e^{i\boldsymbol{q}\cdot\left(\boldsymbol{r}_{p}-\boldsymbol{r}_{p'}\right)}$
for a photonic ball.
To do so, we consider that a PB is a spherical
portion cut-out from an infinite distribution of nanoparticles.
We define a shape function
\begin{equation}
V_{R,\boldsymbol{r}_{c}}\left(\boldsymbol{r}\right)\equiv\begin{cases}
1 & \text{if }\left|\boldsymbol{r}-\boldsymbol{r}_{c}\right|\leq R\\
0 & \text{otherwise}
\end{cases}\label{eq:100}
\end{equation}
that defines the spherical cut. The two point probability density for pairs of nanoparticles in the
infinite system is 
\begin{equation}
\rho^{\left(2\right)}\left(\boldsymbol{r},\boldsymbol{r'}\right)=\rho^{2}g\left(\boldsymbol{r},\boldsymbol{r}'\right)\text{,}\label{eq:110}
\end{equation}
where $\rho$ is the average density of NPs and and $g\left(\boldsymbol{r},\boldsymbol{r}'\right)$
is the radial function that depends only on the difference $\Delta\boldsymbol{r}\equiv\boldsymbol{r}-\boldsymbol{r}'$.

With this we have
\begin{align}
\left\langle \sum_{p,p'}e^{i\boldsymbol{q}\cdot\left(\boldsymbol{r}_{p}-\boldsymbol{r}_{p'}\right)}\right\rangle  & =\int d^{3}\boldsymbol{r}d^{3}\boldsymbol{r}'e^{i\boldsymbol{q}\cdot\left(\boldsymbol{r}-\boldsymbol{r}'\right)}\rho^{\left(2\right)}\left(\boldsymbol{r},\boldsymbol{r'}\right)V_{R,\boldsymbol{r}_{c}}\left(\boldsymbol{r}\right)V_{R,\boldsymbol{r}_{c}}\left(\boldsymbol{r}'\right)\nonumber \\
 & =\rho V_\text{PB}\int d^{3}\left(\Delta\boldsymbol{r}\right)e^{i\boldsymbol{q}\cdot\Delta\boldsymbol{r}}\rho g\left(\Delta\boldsymbol{r}\right)h\left(\Delta\boldsymbol{r},R\right)\label{eq:120}
\end{align}
where $V_{R,\boldsymbol{r}_{c}}\left(\boldsymbol{r}\right)h\left(\Delta\boldsymbol{r},R\right)\equiv\left\langle V_{R,\boldsymbol{r}_{c}}\left(\boldsymbol{r}\right)V_{R,\boldsymbol{r}_{c}}\left(\boldsymbol{r}'\right)\right\rangle $
and 
\begin{equation}
h\left(\Delta\boldsymbol{r},R\right)=\begin{cases}
\left(1+\frac{\Delta r}{4R}\right)\left(1-\frac{\Delta r}{2R}\right)^{2} & \text{if }\Delta r<2R\\
0 & \text{otherwise}
\end{cases}\label{eq:130}
\end{equation}
Our result for the structure factor of
a photonic ball is 
\begin{equation}
 S_\text{PB} (\boldsymbol{q})=\int d^{3}\left(\Delta\boldsymbol{r}\right)e^{i\boldsymbol{q}\cdot\Delta\boldsymbol{r}}\rho g\left(\Delta\boldsymbol{r}\right)h\left(\Delta\boldsymbol{r},R_\text{PB}\right)\label{eq:140}
\end{equation}
Taking into account that 
\begin{equation}
\int d^{3}\Delta\boldsymbol{r}e^{i\boldsymbol{q}\cdot\Delta\boldsymbol{r}}\rho g\left(\Delta r\right)=S\left(\boldsymbol{q}\right)\label{eq:170}
\end{equation}
and 
\begin{equation}
\int d^{3}\left(\Delta\boldsymbol{r}\right)e^{i\boldsymbol{q}\cdot\Delta\boldsymbol{r}}\rho h\left(\Delta r,R_\text{PB}\right)=NP_\text{PB}\left(q, R_\text{PB}\right)\label{eq:180}
\end{equation}
we can rewrite Eq.~(\ref{eq:140}) in an equivalent but more compact form as a convolution of infinite system structure factor $S(\boldsymbol{q})$ and photonic ball RGD form factor $P\left(\boldsymbol{q},  R_\text{PB}\right)$
\begin{equation}
 S_\text{PB} (\boldsymbol{q}) =V_\text{PB} \int S(\boldsymbol{q'})   P\left(\boldsymbol{q}-\boldsymbol{q'},  R_\text{PB}\right) d \boldsymbol{q'}. \label{Eq:PB_convolution}
\end{equation}
Let us recall here that the photonic ball form factor $P\left(\boldsymbol{q},  R_\text{PB}\right)$ is defined so that $P(0,R_\text{PB})=1$. Taking advantage of spherical symmetry of our system, we can rotate $\boldsymbol{q}$ to have only $q_{z}$ component and pass to spherical coordinates. Then
\begin{equation}
 S_\text{PB} (q) =2 \pi V_\text{PB} \iint S(q')   P\left(\sqrt{q^{2}+q'^{2}-2q q' cos\theta},  R_\text{PB}\right) (q')^{2} \sin \theta d q' d\theta. \label{Eq:PB_convolutionSphericalCoordinates}
\end{equation}
So far this expression is essentially exact within the first order
scattering approach. 

\paragraph{Simple analytical model for $S_\text{PB}(q)$}
 Eq.~\eqref{Eq:PB_convolution1}, or equivalently Eq.~(\ref{Eq:PB_convolution}),  can be used to calculate the PB structure factor numerically but it does not reveal much physical insight concerning the different contributions to the scattering signal. Thus, we are also interested in a more simple model which could give simple analytical predictions that asymptotically approach Eq.~(\ref{Eq:PB_convolution}) in both the low and large $q-$ limits.
Starting from 
\begin{equation}
gh=\left(g-1\right)\left(h-1\right)+g+h-1\label{eq:150}
\end{equation}
and taking into account that 
\begin{align}
    g\left(\Delta\boldsymbol{r}\right)=0 &\text{ for } \Delta r<2R_\text{NP}\\
    h\left(\Delta\boldsymbol{r},R_\text{PB}\right)\simeq1 &\text{ for }\Delta r\ll R_\text{PB}\\
    g\left(\Delta\boldsymbol{r}\right)\simeq1 &\text{ for }\Delta r\gg R_\text{NP},
\end{align}
we see that if the radius of the PB is substantially larger than the
radius of a single NP then $\left(g\left(\Delta r\right)-1\right)\left(h\left(\Delta r\right)-1\right)\simeq0$, hence
\begin{equation}
 S_\text{PB} (\boldsymbol{q}) \simeq \int d^{3}\left(\Delta\boldsymbol{r}\right)e^{i\boldsymbol{q}\cdot\Delta\boldsymbol{r}}\rho g\left(\Delta r\right)+\int d^{3}\left(\Delta\boldsymbol{r}\right)e^{i\boldsymbol{q}\cdot\Delta\boldsymbol{r}}\rho h\left(\Delta r\right) \label{eq:160}
\end{equation}
where we dropped the last term in Eq.~(\ref{eq:150}) since it is only
relevant for $q=0$. 
We can now write 
 \begin{equation}
  S_\text{PB} (q) \approx  S\left(q, R_\text{NP}\right )+N P(q,R_\text{PB}). \label{Eq:PBCSA}
\end{equation}
Equation \eqref{Eq:PBCSA} is a simplified version of the analytical model (Eq.~\eqref{Eq:PB_convolution}).
Figure~\ref{fig:AnalyticalModel} shows that this approximation agrees pretty well with the precise model calculated numerically with Eq.~(\ref{Eq:PB_convolution}) for both the low angles ($q R_\text{RB}\to 0$) and high angle ($q R_\text{RB} \gg 1$) limits.
However, it still deviates in the medium $q$ range, which corresponds to distances comparable with both radii $R_\text{PB}$ and $R_\text{NP}$.
This approximation demonstrates that the scattering of a PB can be understood as the algebraic sum of the scattering from the internal structure and the scattering of the entire homogeneous PB.

Using the RGD-form factor for nanoparticles, in the limit $q R_\text{RB}\to 0$,  we find 
\begin{equation}
\frac{d\sigma}{d\Omega}_\text{PB}\left(q R_\text{PB}\right)\propto N^2 \left|\frac{m^2-1}{m^2+2}\right|^2  V_\text{NP}^2 P(q,R_\text{PB}), 
\end{equation}
with $V_\text{NP}$ being nanoparticle volume.  Thus $\frac{d\sigma}{d\Omega}_\text{PB}\left(q R_\text{PB}\to 0\right)$ is equal to the differential scattering cross section of a homogeneous sphere with $R_\text{PB}$ and having a reduced effective index $n_\text{eff}$.
The effective index is set by
\begin{align}
N^2 \left|\frac{m^2-1}{m^2+2}\right|^2  V_\text{NP}^2 \equiv \left|\frac{m_\text{eff}^2-1}{m_\text{eff}^2+2}\right|^2  V_\text{PB}^2,
\end{align}
with $V_\text{PB}=N 
V_\text{NP}/\varphi$.
Thus we naturally obtain that, in the 1\textsuperscript{st} Born approximation, the effective refractive index of a photonic ball corresponds to the Maxwell-Garnett effective medium approximation~\cite{Garnett1904}
\begin{align}
\left|\frac{m^2-1}{m^2+2}\right|  \varphi \equiv \left|\frac{m_\text{eff}^2-1}{m_\text{eff}^2+2}\right| .
\end{align}
For polystyrene NPs ($n=1.59$) in water ($n=1.33$) and for a filling fraction $\varphi=0.6$, this yields $n_\text{eff}=1.483$, in agreement to the results presented in Fig.~\ref{fig:hSphereRGD}~(a).

In the limit $q R_\text{RB} \gg 1$, we naturally retrieve the scattering of an infinite colloidal glass which only depends on the size and shape of the PB through the aggregation number $N$
\begin{equation}
\frac{d\sigma}{d\Omega}_\text{PB}\left(q R_\text{PB}\right)\propto N S\left(q, R_\text{NP}\right )  \frac{d\sigma}{d\Omega}_\text{NP}\left(q,R_\text{NP} \right).
\end{equation}


\newpage
\section*{Supplemental Figures}
 \begin{figure*}[h]
    \centering
    \includegraphics[width=.6\linewidth]{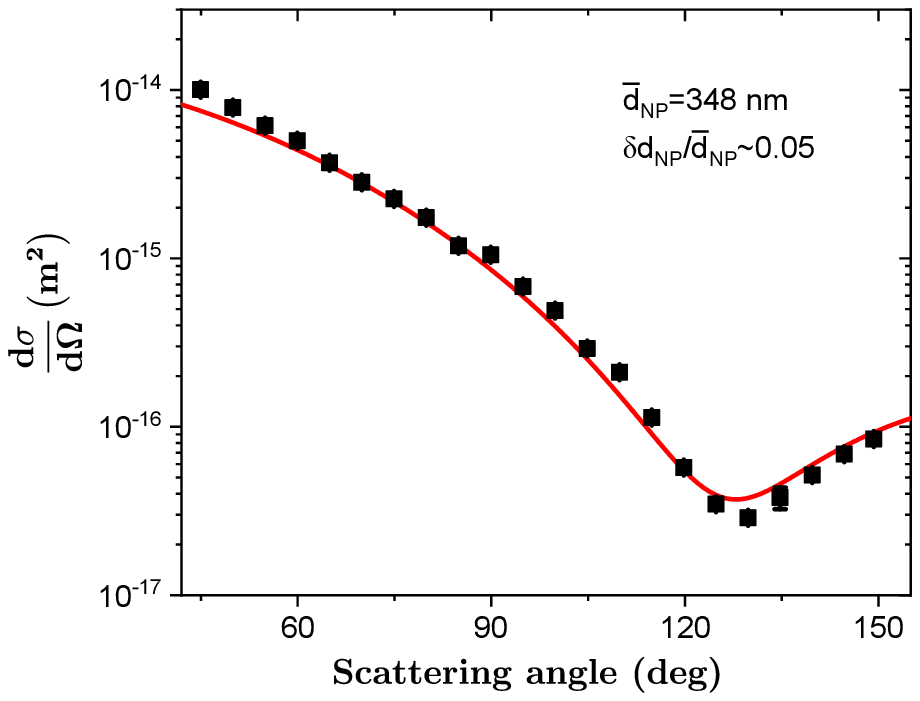}
    \caption{Differential scattering cross-section of nanoparticles measured by static light scattering of dilute suspensions and its comparison to Lorentz-Mie theory.}
    \label{fig:SLS348nm}
\end{figure*}

\begin{figure*}
    \centering
    \includegraphics[width=.6\linewidth]{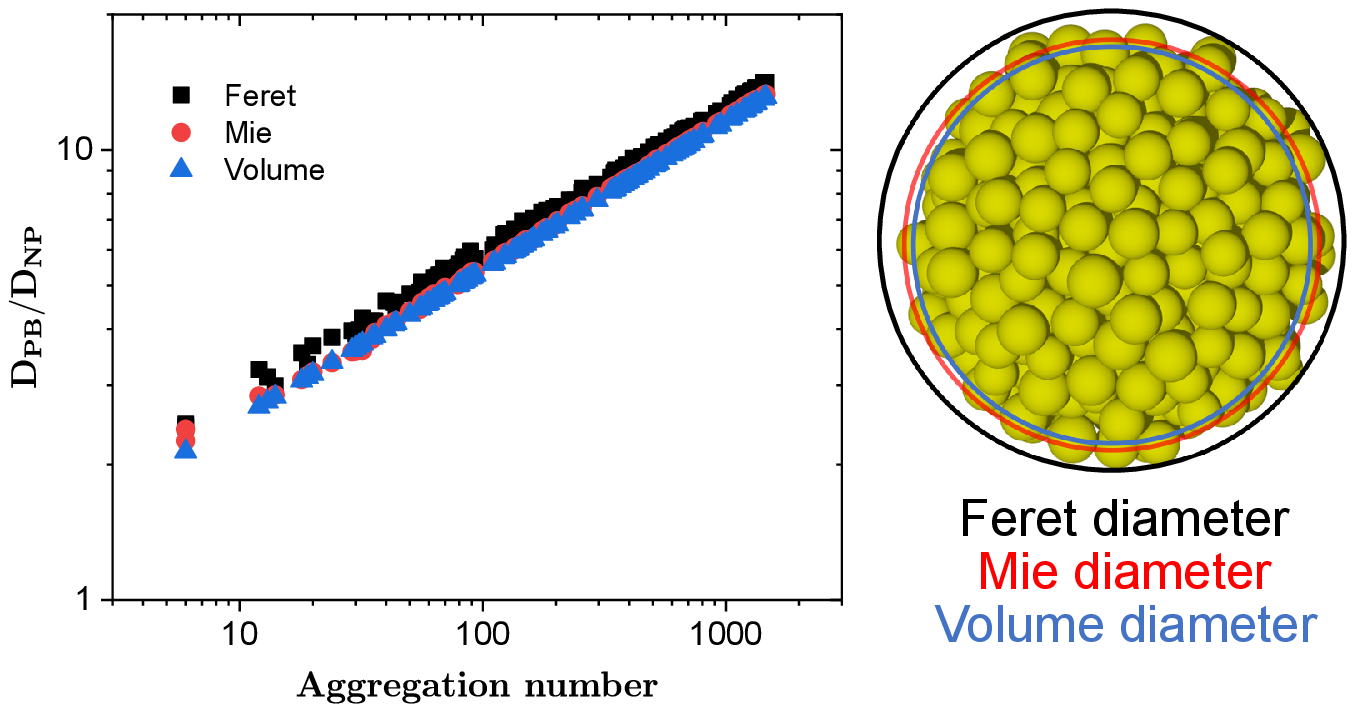}
    \caption{Normalised Diameter of PB using various diameter definitions as a function of PB aggregation number as extracted from simulations. Feret diameter is defined as a diameter of sphere circumscribed around the PB. Feret diameter is of particular importance, as it is extracted from SEM images of PBs. Even if the exact theoretical relationship between Feret diameter and N is not clear, we can use the interpolation of this graph to get the aggregation number N from SEM images. Mie diameter is extracted by fitting the PB differential scattering cross section by homogeneous Lorentz-Mie  sphere cross-section. Volume diameter is defined as $(D_\text{PB}/D_\text{NP})=(N/\varphi)^{1/3}$. One can see that there exists a small but stable difference between all three definitions. b) Schema representing the difference between the various diameter definitions.}
    \label{fig:Diameter}
\end{figure*}
 \begin{figure*}
    \centering
    \includegraphics[width=.6\linewidth]{PBinside.eps}
    \caption{The electron micrographs of potonic balls after ultrasonication reveal the disordered internal structure.
    }
    \label{fig:PBinside}
\end{figure*}
    
   \begin{figure}[h!]
\centering
      \includegraphics[width=.6\linewidth]{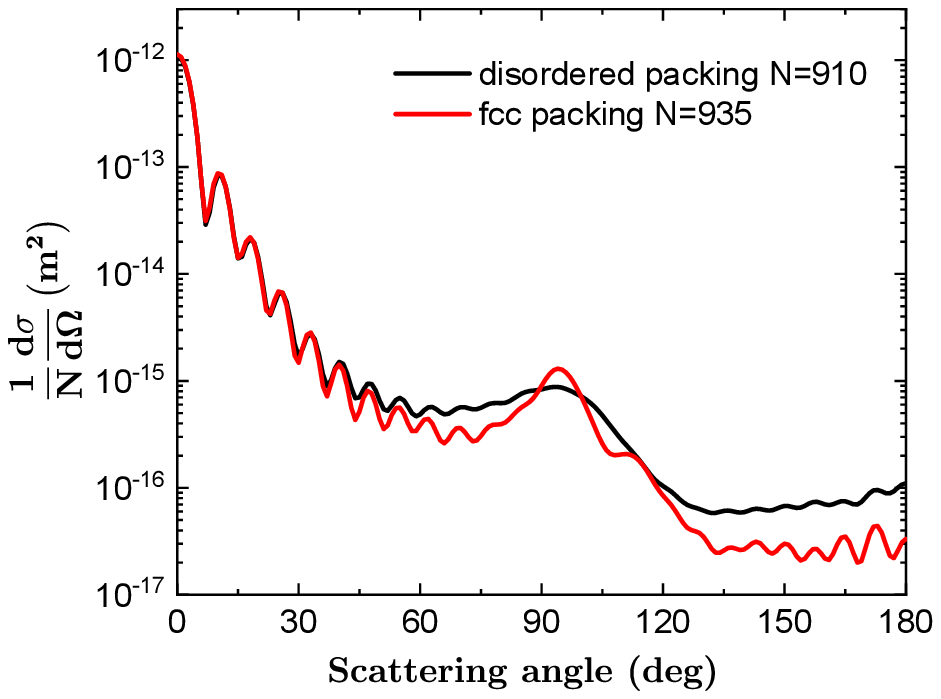}
     \caption{Normalized differential scattering cross-section for disordered and crystallized PBs.}\label{fig:CryDiso}
    \end{figure} 
\begin{figure*}
    \centering
    \includegraphics[width=0.6\linewidth]{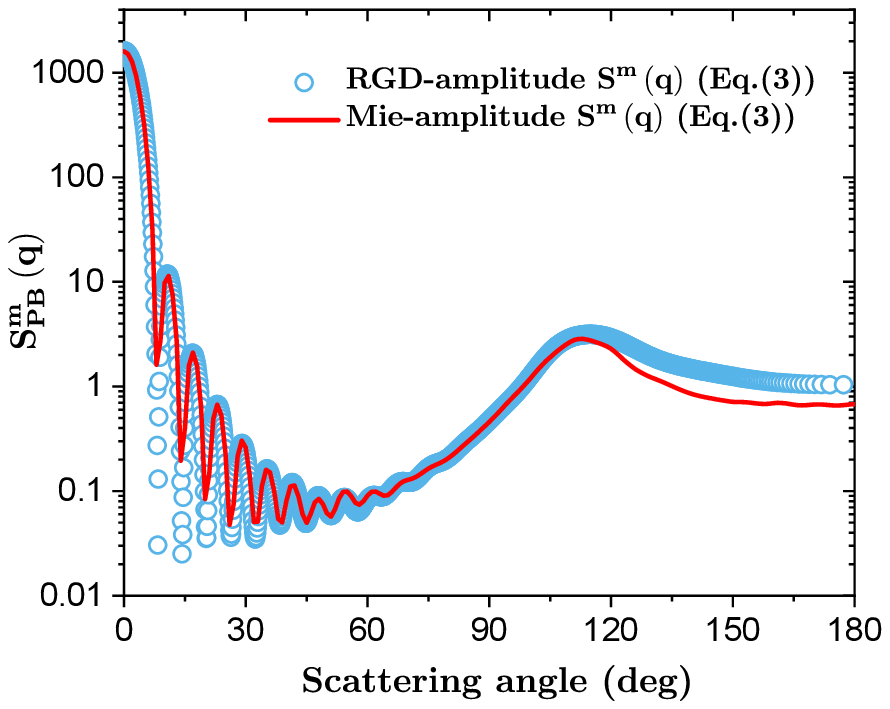}
    \caption{Photonic ball measurable structure factor for $N=1650$, polydispersity 5\%, calculated by the direct calculation from a simulated configuration, as shown in Fig. \ref{fig:CSPB}, using  Eq.~(\ref{eq:PBstructurefactorpolydisperse}) with RGD (circles) and Mie (solid line) scattering amplitudes  of the NPs.}
    \label{fig:MieAmplitudes}
\end{figure*}

\begin{figure*}
\centering
    \includegraphics[width=0.6\linewidth]{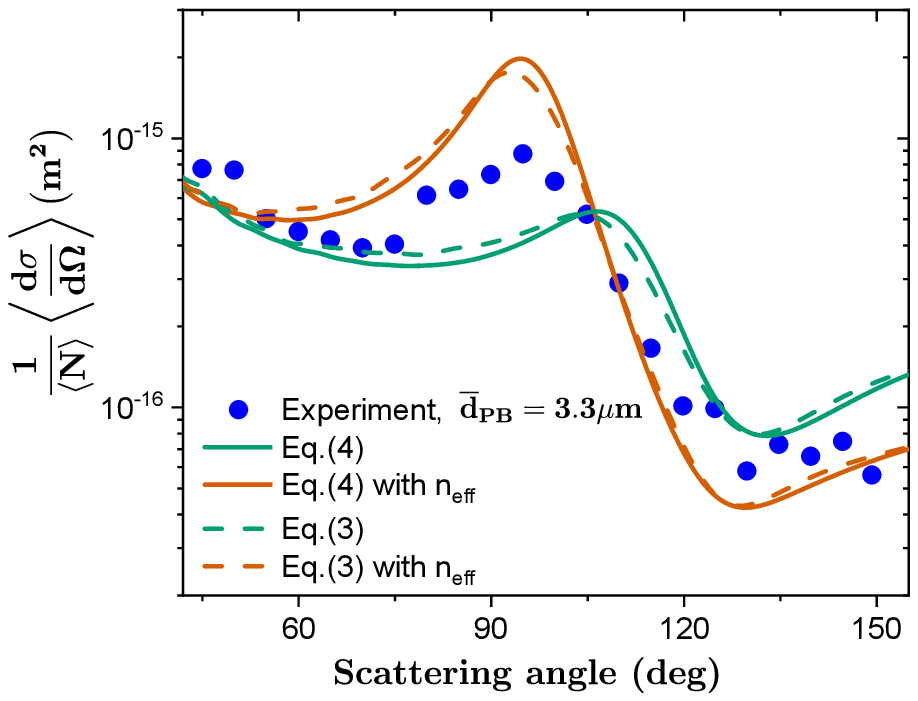}
        \caption{Comparison of experimental differential scattering cross-section of PB dispersion with a mean diameter  $\overline d_\text{PB}=3.3 \mu$m and polydispersity $\delta d_\text{PB}/\overline d_\text{PB} \sim 0.45$ (blue circles), the convolution model Eq.~\eqref{Eq:PB_convolution1} and  numerical calculation Eq.~\eqref{eq:PBstructurefactorpolydisperse}.
        }\label{fig:PBanaSI}
    \end{figure*}

\begin{figure*}
    \centering
    \includegraphics[width=0.6\linewidth]{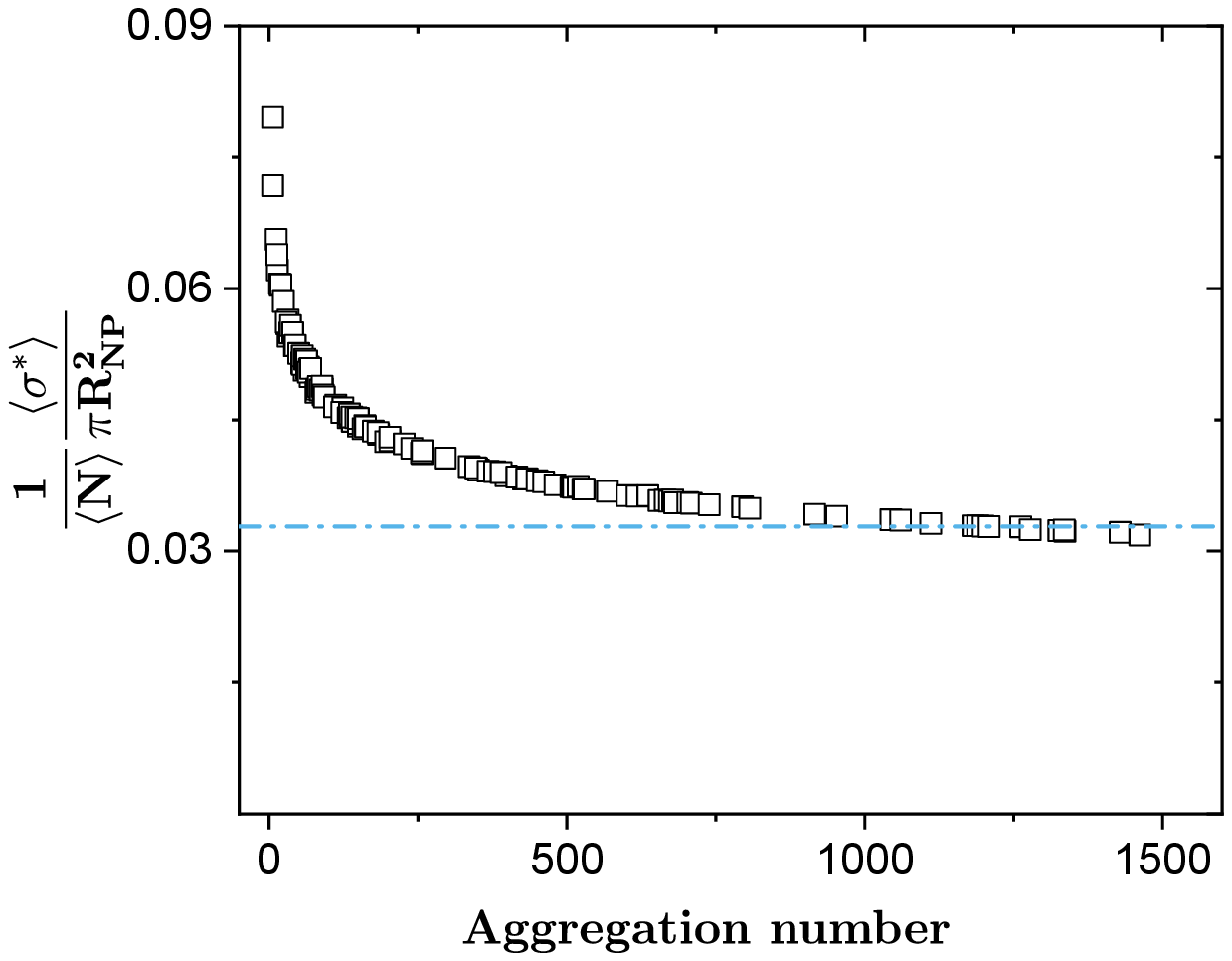}
    \caption{Transport cross-section of PBs in water for different aggregation numbers and fixed wavelength $\lambda / R_{NP}=660/174=3.79$. The blue dashed line represents the prediction for an infinite photonic glass based on the measurable PY structure factor, a Mie NP form factor and the effective refractive index correction (Maxwell Garnett).}
    \label{fig:TransportVSsize}
\end{figure*}

\begin{figure*}
\centering
\includegraphics[width=0.5\columnwidth]{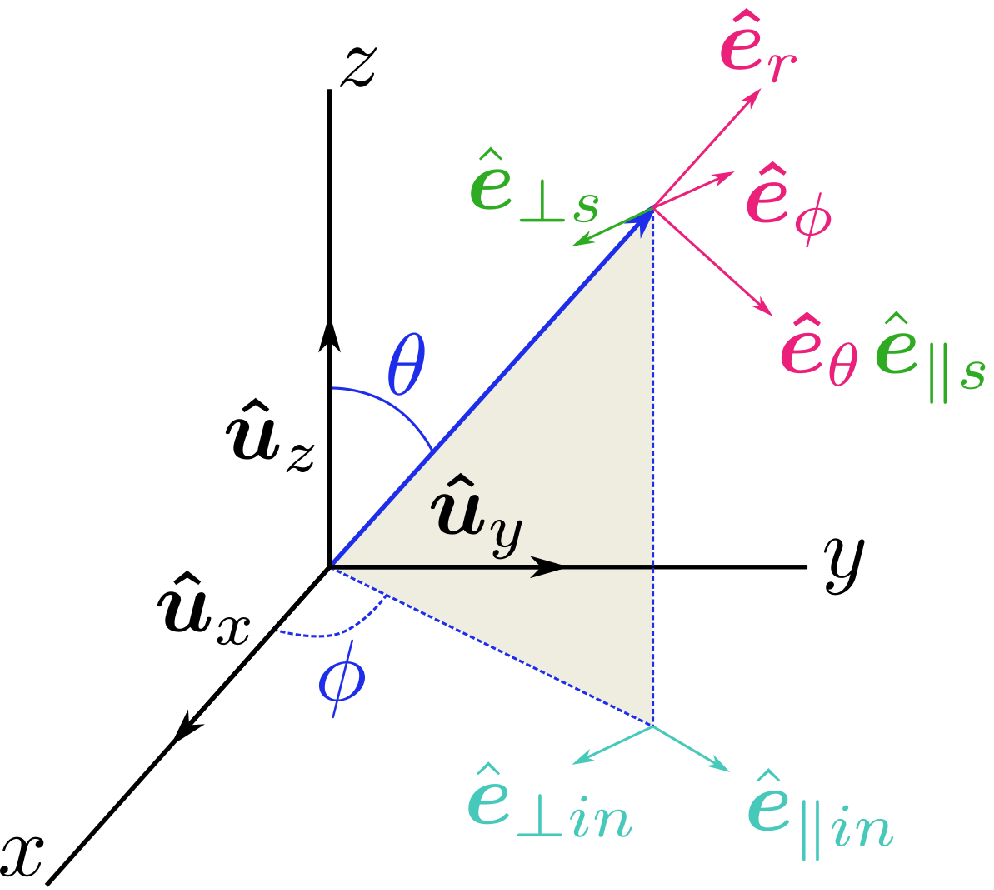}
\caption{\label{fig:sketch_coords}Coordinate system used to describe scattering.}
\end{figure*}

\begin{figure*}
    \centering
    \includegraphics[width=0.6\linewidth]{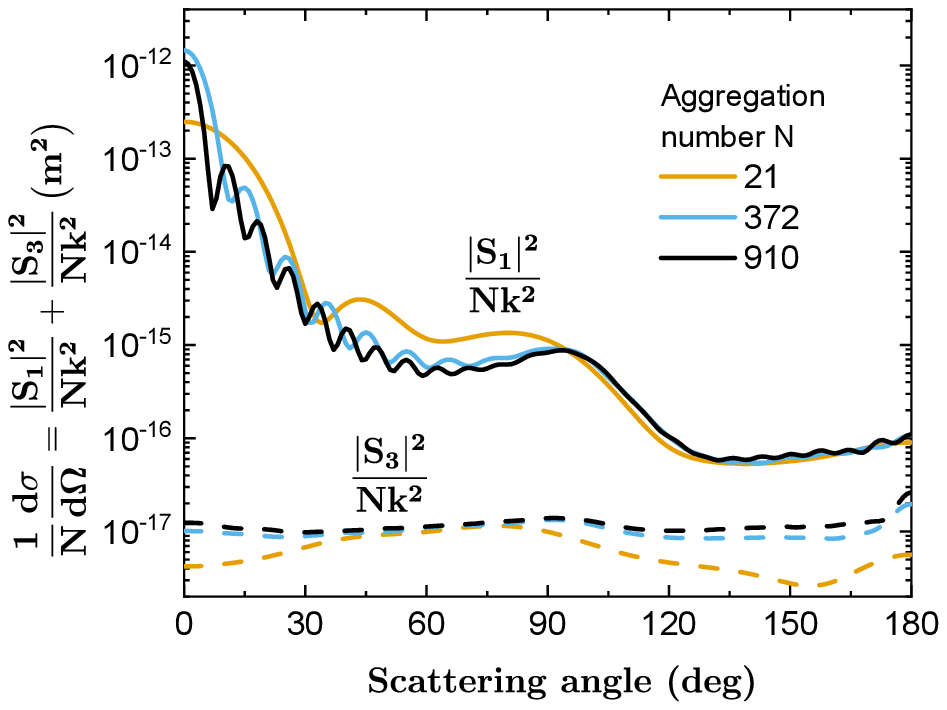}
    \caption{Elements of scattering matrix  for differently sized PBs - $\frac{\lvert S_{1}\rvert^{2}}{Nk^{2}}$ and $\frac{\lvert S_{3}\rvert^{2}}{Nk^{2}}$.}
    \label{fig:MieAmplitudes2}
\end{figure*}

\end{document}